\documentstyle[11pt,epsf]{article}
\def\today{\number\day\space
     \ifcase\month\or
       January\or February\or March\or April\or May\or June\or
       July\or August\or September\or October\or November\or December\fi
     \space\number\year}
\newcommand{\PL}[3]{{Phys. Lett.}        {#1} {(19#2)} {#3}}
\newcommand{\PRL}[3]{{Phys. Rev. Lett.} {#1} {(19#2)} {#3}}
\newcommand{\PR}[3]{{Phys. Rev.}        {#1} {(19#2)} {#3}}
\newcommand{\NP}[3]{{Nucl. Phys.}        {#1} {(19#2)} {#3}}

\newcommand{\beq}{\begin{equation}}
\newcommand{\eeq}{\end{equation}}
\newcommand{\ba}{\begin{array}}
\newcommand{\ea}{\end{array}}
\newcommand{\beqa}{\begin{eqnarray}}
\newcommand{\eeqa}{\end{eqnarray}}
\newcommand{\beqan}{\begin{eqnarray*}}
\newcommand{\eeqan}{\end{eqnarray*}}
\newcommand{\dis}{\displaystyle}
\newcommand{\bit}{\begin{itemize}}
\newcommand{\eit}{\end{itemize}}
\newcommand{\ben}{\begin{enumerate}}
\newcommand{\een}{\end{enumerate}}

\newcommand{\cL}{{\cal L}}

\newcommand{\dg}{\dagger}
\newcommand{\dfrac}{\displaystyle \frac}
\newcommand{\no}{\nonumber}
\newcommand{\rms}{\rm\scriptsize}
\newcommand{\ve}{\varepsilon}
\newcommand{\vp}{\varphi}

\newcommand{\wt}{\widetilde}
\newcommand{\wh}{\widehat}

\newcommand{\lsim}{\stackrel{<}{_\sim}}

\newcommand{\ra}{\rightarrow}

\newcommand{\Imm}{\mbox{Im}}

\begin{document}
\begin{titlepage}
\begin{flushright}
INFNNA-IV-94/24\\
DSF-94/24\\
UWThPh-1994-20\\
CERN-TH-7503/94\\
[1cm]
\end{flushright}
\begin{center}

\vspace*{1cm}

{\Large \bf RADIATIVE NON--LEPTONIC KAON DECAYS*}

\vspace*{2cm}
{\bf{G. D'Ambrosio$^{1 \dagger}$, G. Ecker$^2$, G. Isidori$^3$ and H.
Neufeld$^2$}}

\vspace{1cm}
${}^{1)}$ Theoretical Physics Division, CERN,\\
CH-1211 Geneva 23, Switzerland\\[10pt]

${}^{2)}$ Institut f\"ur Theoretische Physik, Universit\"at Wien\\
A--1090 Wien, Austria \\[10pt]

${}^{3)}$ Istituto Nazionale di Fisica Nucleare, Sezione di Roma \\
Dipartimento di Fisica, Universit\`a di Roma ``La Sapienza''\\
I--00185 Roma, Italy

\end{center}

\vfill
\noindent * Work supported in part
by HCM, EEC--Contract No. CHRX--CT920026 (EURODA$\Phi$NE) and
by FWF (Austria), Project Nos. P09505--PHY, P08485-TEC.

\noindent $\dagger$ On leave of absence from
Istituto Nazionale di Fisica Nucleare, Sezione di Napoli, Naples, Italy.

\vspace*{1cm} \noindent
INFNNA-IV-94/24\\
DSF-94/24\\
UWThPh-1994-20\\
CERN-TH-7503/94\\
November 1994\\

\end{titlepage}
\tableofcontents

\renewcommand{\thesection}{\arabic{section}}
\renewcommand{\thesubsection}{\arabic{section}.\arabic{subsection}}
\renewcommand{\theequation}{\arabic{section}.\arabic{equation}}

\setcounter{equation}{0}
\setcounter{subsection}{0}

\section{Non--leptonic weak interactions}
\label{sec:intro}

\subsection{Introduction}
Many radiative non--leptonic kaon decays will be interesting by--products
of the experimental program at DA$\Phi$NE. The following survey serves
several purposes:
\bit
\item We investigate to what extent DA$\Phi$NE  will be able to test the
Standard Model in the confinement regime with radiative kaon
decays. We concentrate on processes which can be detected at
DA$\Phi$NE and we review briefly those decays where only upper limits
can be expected \footnote{There are several recent reviews of
rare $K$ decays \cite{reviews} which can be consulted for additional
information.}.
\item With reliable predictions from the Standard Model at our disposal,
one can set about looking for new physics. This applies especially to
transitions that are either suppressed or forbidden in the Standard Model.
\item Unambiguous predictions of the Standard Model can be made for the
low--energy structure of non--leptonic weak amplitudes. Chiral perturbation
theory (CHPT) allows us to specify this low--energy structure in terms of some
a priori undetermined low--energy constants. All specific models for
the non--leptonic weak transitions  have to satisfy those
low--energy theorems, but they can be expected to produce
more detailed predictions for the low--energy constants.
\eit

We collect the decays of interest for DA$\Phi$NE in Table~\ref{tab:datatab}.
We have used the average experimental rates from Ref.~\cite{PDG}.
Some of the theoretical estimates are
based on model assumptions that go beyond pure CHPT.
More details on theory and experiments will be found in
the relative sections. To illustrate the improvements at DA$\Phi$NE, we
have assumed the following numbers of tagged events per year
corresponding to a luminosity of $5\cdot 10^{32}~s^{-1}cm^{-2}$ and an
effective year of $10^7 ~s$ \cite{Franzini}:
\beq
 1.1 (1.7) \cdot 10^9 ~K_L (K_S)~, \quad \qquad
9\cdot 10^9 ~K^{\pm}~.
\eeq
The fluxes for $K_L$ decays have to be multiplied for the fiducial volume
of the apparatus.
Furthermore, for channels with experimental limits only or with poor
statistics, we have used the theoretical predictions
to estimate the number of events expected at DA$\Phi$NE.

We have classified the radiative non--leptonic decays in
Table~\ref{tab:datatab} in the following groups:
\bit
\item[$i.$] Two photons in the final state;
\item[$ii.$]  One photon in the final state,
distinguishing between internal bremsstrahlung and direct emission;
\item[$iii.$]  Decays with a lepton pair.
\eit

We also discuss some possible bounds that can be put on CP violating
quantities with radiative non--leptonic decays at DA$\Phi$NE.

\begin{table}
\label{tab:datatab}
\[ \begin{array}{|l|c|c|c|} \hline
\mbox{channel} & BR_{\rm exp}   & BR_{\rm theor}&\mbox{\# {\rm events/yr}}
\\ \hline
      \multicolumn{4}{c}{\rm Two~photons~in~the~final~state.}\\ \hline
K_S\rightarrow\gamma\gamma&(2.4\pm 1.2)\cdot10^{-6}&2.1\cdot10^{-6}&3.6
\cdot 10^3 \\ \hline
K_L\rightarrow\gamma\gamma&(5.73\pm 0.27)\cdot10^{-4}&\sim 5\cdot 10^{-4}
&6.3\cdot10^5  \\ \hline
K_L\rightarrow\pi ^0 \gamma\gamma&(1.70\pm 0.28)\cdot10^{-6}&\sim 10^{-6}
 &1.9 \cdot 10^3 \\ \hline
K_S\rightarrow\pi ^0 \gamma\gamma \hfill _{(M_{\gamma\gamma}>220 MeV)}
&-&3.8\cdot10^{-8}&65\\ \hline
K^+\rightarrow\pi ^+ \gamma\gamma& < 1.5 \cdot 10^{-4}&\sim
5\cdot10^{-7}&\sim 4.5\cdot 10^3\\ \hline
K_L\rightarrow\pi ^0 \pi ^0 \gamma\gamma \qquad \hfill _{(|M_{\gamma\gamma}-
M_\pi|>20 MeV)}
&-&3\cdot 10^{-8}& 33\\ \hline
K_S\rightarrow\pi ^0 \pi ^0 \gamma\gamma \hfill _{(|M_{\gamma\gamma}-
M_\pi|>20 MeV)}
&-&5\cdot 10^{-9}& 8\\ \hline
      \multicolumn{4}{c}{\rm One~photon~in~the~final~state,~internal~
       bremsstrahlung.}\\ \hline
K_S\rightarrow\pi^{+}\pi^{-}\gamma \hfill _{(E^*_{\gamma}>50 MeV)}
 &(1.78\pm0.05)\cdot10^{-3}&1.75\cdot 10^{-3}&3\cdot10^{6}\\ \hline
K_L\rightarrow\pi^{+}\pi^{-}\gamma \hfill _{(E^*_{\gamma}>20 MeV)}
 &(1.49\pm0.08)\cdot10^{-5}&1.42\cdot 10^{-5}&1.5\cdot10^{4}\\ \hline
K^+\rightarrow\pi^+\pi^{0}\gamma \hfill \quad _{(
 T_c^*=(55-90)MeV)}
 &(2.57\pm0.16)\cdot 10^{-4}&2.61\cdot10^{-4}&2.3\cdot10^{6}\\ \hline
         \multicolumn{4}{c}{\rm One~photon~in~the~final~state,~
          direct~emission.}\\ \hline
K_S\rightarrow\pi^{+}\pi^{-}\gamma \hfill _{(
 E^*_\gamma >50MeV)} & <9\cdot 10^{-5}&\sim 10^{-6}
 &\sim 1.7\cdot 10^3 \\ \hline
K_L\rightarrow\pi^{+}\pi^{-}\gamma \hfill _{(E^*_{\gamma}>20 MeV)}
 &(3.19\pm0.16)\cdot 10^{-5}&\sim 10^{-5}&3.5\cdot10^{4}\\ \hline
K^+\rightarrow\pi^+\pi^{0}\gamma  \hfill \quad _{(
 T_c^*=(55-90)MeV)}
 &(1.8\pm 0.4)\cdot 10^{-5}& \sim 10^{-5}&1.6\cdot10^{5}\\ \hline
 \multicolumn{4}{c}{\rm Lepton~pair~without~pions.} \\ \hline
K_L\rightarrow \mu^+ \mu^- &(7.4\pm 0.4)\cdot 10^{-9} & \sim 7\cdot 10^{-9}
& 8 \\ \hline
K_L\rightarrow\gamma e^+e^-&(9.1\pm0.5)\cdot 10^{-6}
 &9\cdot 10^{-6}&1.0\cdot 10^4\\ \hline
K_L\rightarrow\gamma \mu^+\mu^-&(2.8\pm2.8)\cdot 10^{-7}&3.6\cdot 10^{-7}
&4.0\cdot 10^2\\  \hline
K_S\rightarrow\gamma e^+e^-&-&3.4\cdot 10^{-8}&58\\ \hline
K_L\rightarrow e^+e^-e^+e^-&(3.9\pm 0.7)\cdot 10^{-8}&-&43\\ \hline
          \multicolumn{4}{c}{\rm Lepton~pair~with~pions.}\\ \hline
K_S\rightarrow\pi^{0}e^{+}e^{-}&<1.1\cdot 10^{-6}
 &>5\cdot10^{-10}&> 1\\ \hline
K_S\rightarrow\pi^{0}\mu^{+}\mu^{-}&-&> 10^{-10}&-\\ \hline
K^+\rightarrow\pi^+ e^{+}e^{-}&(2.74\pm0.23)\cdot 10^{-7}&
\sim 3\cdot 10^{-7}& 2.5\cdot 10^3\\ \hline
K^+\rightarrow\pi^+\mu^{+}\mu^{-}&<2.3\cdot10^{-7}
& 6\cdot10^{-8}& 5.4\cdot 10^2\\ \hline
K_L\rightarrow \pi^+\pi^-e^+e^- &< 2.5 \cdot 10^{-6}&2.8\cdot10^{-7}
&3.1 \cdot 10^2 \\ \hline
\end{array}\]
\caption{Radiative kaon decays of interest for DA$\Phi$NE.}
\end{table}

\subsection{Chiral perturbation theory}
\label{subsec:CHPT}
CHPT for the non--leptonic weak
interactions is a straightforward extension of the standard
CHPT formalism for the strong, electromagnetic and semileptonic
weak interactions (for a general introduction and additional references
we refer to \cite{BEG}). The $\Delta S=1$ non--leptonic weak interactions
are treated as a perturbation of the strong chiral Lagrangian. We
only consider weak amplitudes to first order in the weak interactions,
$O(G_F)$.

The effective Hamiltonian for $\Delta S=1$ weak interactions at energies
much smaller than $M_W$ takes the form of an operator product expansion
\cite{GL74}
\beq\label{eq:effham}
{\cal H}^{\Delta S = 1}_{\mbox{\rms eff}}\, = \, {G_F \over
\sqrt{2}}  V_{ud} V_{us}^* \,
\sum_{i} C_i \, Q_i \, +
\, \mbox{h.c.}
\eeq
in terms of four--quark operators $Q_i$ and Wilson coefficients $C_i$.
Under the chiral group $SU(3)_L \times SU(3)_R$, the effective Hamiltonian
(\ref{eq:effham}) transforms as
\beq
 {\cal H}^{\Delta S = 1}_{\mbox{\rms eff}}\, \sim \,
(8_L,1_R)\, + \, (27_L,1_R) ~. \label{827}
\eeq

Due to the Goldstone theorem, the effective chiral Lagrangian
for the $\Delta S=1$ non--leptonic weak interactions starts
at $O(p^2)$. The most general chiral Lagrangian of lowest order
with the same transformation properties as (\ref{827}) has the form
\footnote{We use the same conventions as in \cite{BEG}.}
\beq\label{eq:L2w}
\cL_2^{\Delta S=1} =
G_8  \langle \lambda L_\mu L^\mu\rangle +
G_{27} \left( L_{\mu 23} L^\mu_{11} + \dfrac{2}{3} L_{\mu 21}
L^\mu_{13}\right) + \mbox{\rm h.c.}  ,
\eeq
where
\beq
\lambda =  (\lambda_6 - i \lambda_7) /2 ~, \qquad\qquad
L_\mu = i F^2 U^\dg D_\mu U ~,
\eeq
and $\langle A \rangle$ denotes the trace of the matrix $A$.
The chiral couplings $G_8$ and $G_{27}$ measure the strength
of the two parts in the effective Hamiltonian (\ref{eq:effham})
transforming as $(8_L,1_R)$ and $(27_L,1_R)$, respectively,
under chiral rotations. From $K\to 2\pi$ decays one finds:
\beq\label{eq:G8G27}
|G_8| \simeq 9 \cdot 10^{-6}~\mbox{GeV}^{-2}, \qquad\qquad
G_{27} / G_8 \simeq 1/18 ~.\label{eq:G827}
\eeq
The big difference between the two couplings is a manifestation
of the $|\Delta I|=1/2$ rule.

The effective Lagrangian (\ref{eq:L2w}) gives rise to the current algebra
relations between $K\ra 2 \pi$ and $K\ra 3 \pi$ amplitudes. However,
for the radiative transitions under consideration the lowest--order
amplitudes due to (\ref{eq:L2w}) are \lq\lq trivial" in the following
sense:
\ben
\item Non--leptonic $K$ decay amplitudes with any number of real
or virtual photons and with at most one pion in the final state
vanish at $O(p^2)$ \cite{EPRb,EPRc}:
\beq
A(K\ra [\pi] \gamma^* \dots \gamma^*) = 0 \qquad \mbox{ at  } O(p^2)~.
\label{eq:pig*}
\eeq
\item The amplitudes for two pions and any number of real or virtual photons
in the final state factorize at $O(p^2)$ into the on--shell amplitude
for the corresponding $K\ra \pi\pi$ decay and a generalized
bremsstrahlung amplitude independent of the specific decay
\cite{RafRing,ENPa,ENPb}:
\beq
A(K\ra \pi\pi \gamma^* \dots \gamma^*)= A(K\ra \pi\pi) A_{\rm brems}~.
\label{brems}
\eeq
\item A similar statement holds for the decays $K\ra 3 \pi
\gamma$: the amplitude of $O(p^2)$ is completely determined by the
corresponding non--radiative decay $K\ra 3 \pi$ \cite{DEN}.
We will not discuss these decays in any detail here.
\een

Therefore, the non--trivial aspects of radiative non--leptonic kaon decays
appear at $O(p^4)$ only. Similarly to the strong sector,
the non--leptonic weak amplitudes consist in general of several parts:
\ben
\item[i.] Tree--level amplitudes from the most general effective chiral
Lagrangian $\cL_4^{\Delta S=1}$ of $O(p^4)$ with the
transformation properties (\ref{827}).
\item[ii.] One--loop amplitudes from diagrams with a single vertex
of $\cL_2^{\Delta S=1}$ in the loop.
\item[iii.] Reducible tree--level amplitudes with a single vertex from
$\cL_2^{\Delta S=1}$ and with a single vertex from the strong Lagrangian
$\cL_4$ or from the anomaly (cf. \cite{BEG}).
\item[iv.] Reducible one--loop amplitudes, consisting
of a strong loop diagram connected to a vertex of $\cL_2^{\Delta S=1}$
by a single meson line. A typical diagram of this type
contains an external $K-\pi$ or $K-\eta$ transition, possibly
with one or two photons (generalized ``pole diagrams''). The calculation
of such diagrams is simplified by a rediagonalization of the kinetic
and mass terms of $\cL_2 + \cL_2^{\Delta S=1}$ (``weak rotation''
\cite{EPRb,EPRc}).
\een

For the tree--level amplitudes of $O(p^4)$ we shall only consider
the octet part. The corresponding Lagrangian can be written as
\cite{KMWa,EKW}
\beq
\cL_4^{\Delta S=1} = G_8 F^2 \sum_i N_i W_i + {\rm h.c.}
\label{eq:L4w}
\eeq
with dimensionless coupling constants $N_i$ and octet operators $W_i$.
Referring to Ref.~\cite{EKW} for the complete Lagrangian, we list
in Table~\ref{tab:Wi} only the terms relevant for radiative decays.

\begin{table}
\label{tab:Wi}
$$
\begin{tabular}{|r|c|c|} \hline
i  & $ W_i$ & $ Z_i$ \\ \hline
14 &
$i \langle \lambda \{ F_L^{\mu\nu} + U^\dg F_R^{\mu\nu} U,
D_\mu U^\dg D_\nu U\} \rangle$ & $1/4 $ \\
15 &
$i \langle \lambda D_\mu U^\dg (U F_L^{\mu\nu} U^\dg +
F_R^{\mu\nu}) D_\nu U \rangle$ & $1/2$ \\
16 &
$i \langle \lambda \{ F_L^{\mu\nu} - U^\dg F_R^{\mu\nu} U,
D_\mu U^\dg D_\nu U\} \rangle$ & $-1/4$ \\
17 &
$i \langle \lambda D_\mu U^\dg (U F_L^{\mu\nu} U^\dg -
F_R^{\mu\nu}) D_\nu U \rangle $ & 0 \\
18 &
$2 \langle \lambda (F_L^{\mu\nu} U^\dg F_{R\mu\nu} U +
U^\dg F_{R\mu\nu} U F_L^{\mu\nu}) \rangle $ & $ - 1/8$ \\
28 &
$ i \ve_{\mu\nu\rho\sigma} \langle \lambda D^\mu U^\dg U \rangle
\langle U^\dg D^\nu U D^\rho U^\dg D^\sigma U \rangle$ & 0 \\
29 &
$2 \langle \lambda [U^\dg \wt F_R^{\mu\nu} U, D_\mu U^\dg D_\nu U]
\rangle $ & 0 \\
30 &
$ \langle \lambda U^\dg D_\mu U \rangle \langle (\wt F_L^{\mu\nu} +
U^\dg \wt F_R^{\mu\nu} U) D_\nu U^\dg U \rangle$  & 0 \\
31 &
$ \langle \lambda U^\dg D_\mu U \rangle \langle (\wt F_L^{\mu\nu} -
U^\dg \wt F_R^{\mu\nu} U) D_\nu U^\dg U \rangle$  & 0 \\ \hline
\end{tabular}
$$
\caption{The octet operators $W_i$ of the $O(p^4)$ weak chiral
Lagrangian (\protect{\ref{eq:L4w}}) relevant for non--leptonic
radiative $K$ decays, together with the corresponding renormalization
constants $Z_i$. The dual field strength tensors are denoted as
$\wt F_{L,R}^{\mu\nu}$. Otherwise, the notation is the same as in
\protect{\cite{BEG}}.}
\end{table}

The non--leptonic weak loop amplitudes are in general divergent.
As in the strong sector with the corresponding low--energy constants
$L_i$, the weak constants $N_i$ absorb the remaining divergences.
Using dimensional regularization for the loop diagrams,
the $N_i$ are decomposed as
\beqa
N_i & = & N_i^r(\mu) + Z_i \Lambda(\mu)  \label{eq:Zi}\\
\Lambda(\mu) & = & {\mu^{d-4}\over 16 \pi^2} \left\{
{1\over d-4} -{1\over 2} \left[ \ln{(4\pi)} + \Gamma'(1) + 1 \right]
\right\} \no
\eeqa
with an arbitrary scale parameter $\mu$.
The constants $Z_i$ listed in Table~\ref{tab:Wi} are chosen to absorb
the one--loop divergences in the amplitudes \cite{KMWa,EKW,EEF}. The scale
dependences of the coupling constants and of the loop amplitude cancel
in any physical quantity. The final amplitudes of $O(p^4)$ are finite
and scale independent.

The renormalized coupling constants $N_i^r(\mu)$ are measurable quantities.
A crucial question is whether all the coupling constants $N_i$ corresponding
to the operators $W_i$ in Table~\ref{tab:Wi} can be measured in radiative
$K$ decay experiments. We list in Table~\ref{tab:Ni} all the non--leptonic
radiative transitions to which the $N_i$ contribute. There are other
decays not sensitive to the $N_i$
that are either given by finite one--loop amplitudes and/or
anomalous contributions
at $O(p^4)$ ($K_S\ra \gamma^* \gamma^*$, $K^0\ra \pi^0 \gamma \gamma$,
$K^0\ra \pi^0\pi^0 \gamma\gamma$, $K_L\ra \pi^+\pi^-\gamma [\gamma]$)
or which vanish even at $O(p^4)$ ($K_L\ra \gamma^*\gamma^*$,
$K^0\ra \pi^0\pi^0\gamma$).

\begin{table}
\label{tab:Ni}
$$
\begin{tabular}{|c|c|c|c|} \hline
$\pi$ & $2 \pi$ & $3 \pi$ & $N_i$ \\ \hline
$\pi^+ \gamma^*$ & & & $N_{14}^r - N_{15}^r$\\
$\pi^0 \gamma^*~(S)$ & $\pi^0\pi^0\gamma^*~(L)$ & & $2 N_{14}^r + N_{15}^r$\\
$\pi^+ \gamma\gamma$ & $\pi^+\pi^0\gamma\gamma$ & & $N_{14} - N_{15}
 -2 N_{18}$ \\
 & $\pi^+\pi^-\gamma\gamma~(S)$ & & " \\
 & $\pi^+\pi^0\gamma$ & $\pi^+\pi^+\pi^-\gamma$ & $N_{14}-N_{15}-N_{16}
 -N_{17}$ \\
 & $\pi^+ \pi^- \gamma~(S)$ & $\pi^+\pi^0\pi^0\gamma$ & " \\
 & & $\pi^+\pi^-\pi^0\gamma~(L)$ & " \\
 & & $\pi^+\pi^-\pi^0\gamma~(S)$ & $7(N_{14}^r-N_{16}^r)+ 5(N_{15}^r
 + N_{17}^r)$ \\
 & $\pi^+\pi^-\gamma~(L)$ & $\pi^+\pi^-\pi^0\gamma~(S)$ & $N_{29} + N_{31}$ \\
 & & $\pi^+\pi^+\pi^-\gamma$ & " \\
 & $\pi^+\pi^0\gamma$ & $\pi^+\pi^0\pi^0\gamma$ & $3 N_{29} - N_{30}$ \\
 & & $\pi^+\pi^-\pi^0\gamma~(S)$ & $2 (N_{29} + N_{31}) + 3 N_{29} -
N_{30}$ \\
 & & $\pi^+\pi^-\pi^0\gamma~(L)$ & $6 N_{28} - 4 N_{30}+ 3 N_{29} - N_{30}$
\\ \hline
\end{tabular}
$$
\caption{Decay modes to which the coupling constants $N_i$ contribute.
For the $3 \pi$ final states, only the single photon channels are
listed. For the neutral modes, the letters $L$ or $S$ in brackets
distinguish between $K_L$ and $K_S$ initial states in the limit of
CP conservation.}
\end{table}

The information contained in Table~\ref{tab:Ni} leads to the following
conclusions:
\bit
\item Read horizontally, one finds all parameter--free relations
between radiative amplitudes of $O(p^4)$. If in the last column the
renormalized constants $N_i^r$ are displayed, the corresponding
decays have divergent one--loop amplitudes. The other modes have
finite loop amplitudes.
\item Read vertically, we infer that from decays with at most two pions
in the final state only the following combinations of counterterm
coupling constants can in principle be extracted:
\beq
N_{14},\, N_{15},\, N_{16}+N_{17},\, N_{18},\, N_{29}+N_{31},\,
3 N_{29}-N_{30}~.
\eeq
Experiments at DA$\Phi$NE are expected to measure most of those decays.
\item Decays with three pions in the final state are needed to determine
$N_{16}$ and $N_{17}$ separately and the combination $3 N_{28} -
2 N_{30}$.
\item Whereas all \lq\lq electric" constants $N_{14},\dots,N_{18}$
can in principle be determined phenomenologically, this is not
the case for the \lq\lq magnetic" constants (the corresponding
operators $W_i$ contain an $\ve$ tensor): only three combinations of
the four constants $N_{28},\dots,N_{31}$ appear in measurable
amplitudes.
\eit

Except for the radiative decays with three pions in the final state
\cite{DEN}, all amplitudes appearing in Table~\ref{tab:Ni} have
been fully calculated to $O(p^4)$. Since DA$\Phi$NE will probably only
be able to measure the lowest--order bremsstrahlung contributions
for decays of the type $K\ra 3 \pi \gamma$, but not the interesting
$O(p^4)$ parts, the phenomenological determination of the $N_i$ will
remain incomplete. Nevertheless, the information expected from DA$\Phi$NE
will be extremely valuable both for checking the parameter--free
low--energy theorems of $O(p^4)$ between different amplitudes and
for testing the predictions of various models for the coupling constants
$N_i$ \cite{EKW,EPRd,Cheng,BP93,BEP}.

For non--leptonic $K$ decays, the relevant expansion parameter for the
chiral expansion of amplitudes is
\beq
{M_K^2\over (4\pi F_\pi)^2} = 0.18~.
\eeq
Although this expansion parameter is reasonably small,
higher--order corrections beyond $O(p^4)$
may in some cases be sizeable. There is at this time no complete
investigation of such higher--order effects even in the
strong sector, not to speak of the non--leptonic weak sector. Some exploratory
studies have already been performed and we shall come back
to them in the subsequent sections.
Already now, we want to emphasize that those investigations
should be viewed as attempts to locate the dominant higher--order effects
rather than predictions of the same theoretical quality as the leading
$O(p^4)$ amplitudes.

\subsection{The chiral anomaly in the non--leptonic weak sector}
\label{subsec:anomaly}
The contributions of the chiral anomaly to strong, electromagnetic and
semileptonic weak amplitudes can be expressed in terms of the
Wess--Zumino--Witten (WZW) functional \cite{WZW} $Z[U,l,r]_{WZW}$.
Its explicit form can be found in Ref.~\cite{BEG}. However,
the chiral anomaly also contributes to non--leptonic weak amplitudes
starting at $O(p^4)$. Two different manifestations of the anomaly
can be distinguished.

The reducible anomalous amplitudes \cite{ENPa,ENPb} (type iii
in the classification of Sect.~\ref{subsec:CHPT}) arise from the
contraction of meson lines between a weak $\Delta S = 1$ Green function and
the WZW functional. At $O(p^4)$, there can only be one such contraction and
the weak vertex must be due to the lowest--order non--leptonic Lagrangian
$\cL_2^{\Delta S=1}$ in Eq.~(\ref{eq:L2w}).
Since $\cL_2^{\Delta S=1}$ contains bilinear terms in the meson fields,
the so--called pole contributions to anomalous non--leptonic amplitudes
can be given in closed form by a simultaneous diagonalization
\cite{EPRb,EPRc} of the kinetic parts of the Lagrangians $\cL_2$ and
$\cL_2^{\Delta S=1}$. The corresponding local Lagrangian
(octet part only) is \cite{ENPa,ENPb}:
\beq
\cL_{\rm an}^{\Delta S=1} =  \dfrac{ieG_8}{8\pi^2F} \wt F^{\mu\nu}
\partial_\mu \pi^0 K^+ \stackrel{\leftrightarrow}{D_\nu} \pi^- +
\dfrac{\alpha G_8}{6\pi F} \wt F^{\mu\nu} F_{\mu\nu}
\left(K^+ \pi^- \pi^0 - \dfrac{1}{\sqrt{2}} K^0 \pi^+ \pi^-\right) +
{\rm h.c.}
\label{Law}
\eeq
Here
$F_{\mu\nu} = \partial_\mu A_\nu - \partial_\nu A_\mu$ is the
electromagnetic field strength tensor, $\wt F_{\mu\nu} = \ve_{\mu\nu
\rho\sigma} F^{\rho\sigma}$ its dual and
$D_\mu \vp^\pm = (\partial_\mu \pm ieA_\mu)\vp^\pm$ denotes the
covariant derivative with respect to electromagnetism.

There are also other reducible anomalous amplitudes. A generic example
is provided by a non--leptonic Green function where an external
$\pi^0$ or $\eta$ makes an anomalous transition to two photons.
Such transitions are the dominant $O(p^4)$ contributions to the decays
$K_S \ra \pi^0\gamma\gamma$ \cite{EPRb} and $K_L \ra \pi^0\pi^0\gamma\gamma$
\cite{DFR,FK93}. All reducible anomalous amplitudes of $O(p^4)$ are
proportional to $G_8$ in the octet limit. No other unknown parameters
are involved.

The second manifestation of the anomaly in non--leptonic weak amplitudes
arises diagrammatically from the contraction of the $W$ boson field
between a strong Green function on one side and the WZW functional on
the other side. However, such diagrams cannot be taken literally at a
typical hadronic scale because of the presence of strongly interacting
fields on both sides of the $W$. Instead, one must first
integrate out the $W$ together with the heavy quark fields. The
operators appearing in the operator product expansion must then be
realized at the bosonic level in the presence of the anomaly.

\begin{table}
\label{taban}
$$
\begin{tabular}{|l|cccccc|} \hline
Transition & $\cL_{\rm an}^{\Delta S=1}$ & $W_{28}$ & $W_{29}$ & $W_{30}$
& $W_{31}$ & expt. \\ \hline
$K^+ \ra \pi^+ \pi^0 \gamma$       & x &   & x & x &   & x \\
$K^+ \ra \pi^+ \pi^0 \gamma\gamma$ & x &   & x & x &   &    \\
$K_L \ra \pi^+ \pi^- \gamma$       &   &   & x &   & x & x  \\
$K_L \ra \pi^+ \pi^- \gamma\gamma$ & x &   & x &   & x &    \\
$K^+ \ra \pi^+ \pi^0 \pi^0\gamma$  &   &   & x & x &   & x  \\
$K^+ \ra \pi^+ \pi^0 \pi^0 \gamma\gamma$ & & & x & x & & \\
$K^+ \ra \pi^+ \pi^+ \pi^- \gamma$ &   &   & x &   & x & x \\
$K^+ \ra \pi^+ \pi^+ \pi^- \gamma\gamma$ & & & x & & x & \\
$K_L \ra \pi^+ \pi^- \pi^0\gamma$ & & x & x & x && \\
$K_S \ra \pi^+ \pi^- \pi^0\gamma(\gamma)$ & & & x & x & x &  \\ \hline
\end{tabular}
$$
\caption{A complete list of local anomalous non--leptonic weak $K$ decay
amplitudes of $O(p^4)$ in the limit of CP conservation.}
\end{table}

Following the methods of Ref.~\cite{PR91}, the bosonization of four--quark
operators in the odd--intrinsic parity sector was investigated in
Ref.~\cite{BEP}. As in the even--intrinsic parity sector, the bosonized
four--quark operators contain factorizable (leading in $1/N_c$, where
$N_c$ is the number of colours) and non--factorizable parts (non--leading
in $1/N_c$).
Due to the non--renormalization theorem \cite{AB69} of the chiral anomaly,
the factorizable contributions of $O(p^4)$ can be calculated exactly
\cite{BEP}. It turns out that the factorizable contributions produce
all the relevant octet operators
proportional to the $\ve$ tensor ($W_{28}$, $W_{29}$, $W_{30}$ and
$W_{31}$ in Table \ref{tab:Wi}).
The non--factorizable parts automatically have the right octet
transformation property (they do not get any contribution from the anomaly)
and are therefore also of the form $W_{28}, \dots$~, $W_{31}$.
Altogether, the $\Delta S=1$ effective Lagrangian in
the anomalous parity sector of $O(p^4)$ can be characterized by the
coefficients \cite{BEP}
\beq
\ba{ll}
N_{28}^{\rm an} = \dfrac{a_1}{8\pi^2} \qquad \qquad &
N_{29}^{\rm an} = \dfrac{a_2}{32\pi^2} \\[10pt]
N_{30}^{\rm an} = \dfrac{3a_3}{16\pi^2} \qquad \qquad &
N_{31}^{\rm an} = \dfrac{a_4}{16\pi^2} ~,
\ea  \label{Nan}
\eeq
where the dimensionless coefficients $a_i$ are expected to be positive
and of order one (most probably smaller than one \cite{BEP}).

In Table \ref{taban} we list all kinematically allowed non--leptonic
$K$ decays that are sensitive either to the
anomalous Lagrangian $\cL_{\rm an}^{\Delta S=1}$ in
(\ref{Law}) or to the direct terms of $O(p^4)$ via (\ref{Nan}).
A separate column indicates
whether the corresponding decay has been observed experimentally.
The transitions with either three pions and/or two
photons in the final state are in general also subject to non--local
reducible anomalous contributions. In the non--leptonic weak sector,
the chiral anomaly contributes only to {\em radiative} $K$ decays.

\setcounter{equation}{0}
\setcounter{subsection}{0}

\section{Kaon decays with two photons in the final state}
\label{sec:gg}

Two photons can have either $CP=+1$
or $CP=-1$. Thus in the case of only two photons in
the final state, due to gauge invariance, the amplitude will be proportional
either to $F_{\mu\nu}F^{\mu\nu}$ (parallel polarization, $CP=+1$) or to
$\varepsilon_{\mu\nu\rho\lambda}F^{\mu\nu}F^{\rho\lambda}$ (perpendicular
polarization, $CP=-1$). In the case of one or
more pions in the final state also other invariant amplitudes will appear.

\subsection{  $K_S \rightarrow \gamma \gamma$}
\begin{figure}
    \begin{center}
       \setlength{\unitlength}{1truecm}
       \begin{picture}(10.0,7.0)
       \epsfxsize 10. true  cm
       \epsfysize 7.  true cm
       \epsffile{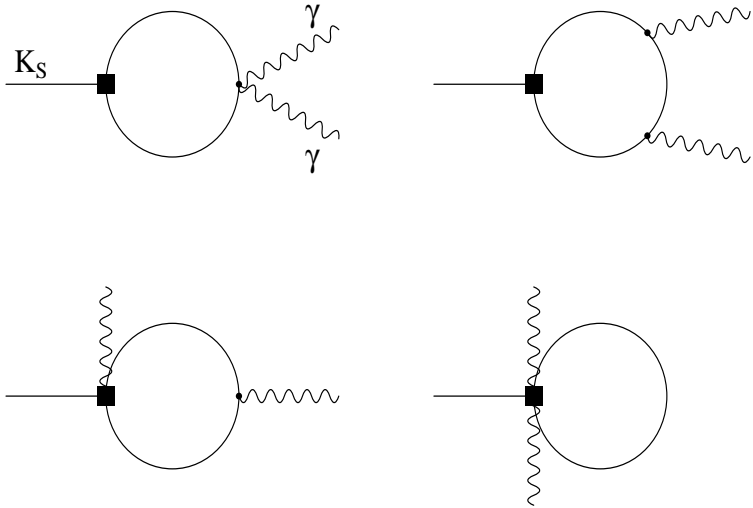}
       \end{picture}
    \end{center}
    \caption{One--loop diagrams for $K_S \to \gamma\gamma$.
	     The black squares indicate the weak vertices.}
    \protect\label{fig:ksloops}
\end{figure}

 We will consider the
CP conserving amplitude $A(K_S \rightarrow \gamma\gamma)$
in the framework of CHPT \cite{DEG}.
Since $K^0$ is neutral, there is no tree--level contribution to
$K^0 \rightarrow \gamma \gamma$. At $O(p^4)$, there are in
principle both chiral meson loops
(Fig.~\ref{fig:ksloops}) and $O(p^4)$ counterterms. Again
because of $K^0$ being neutral, $O(p^4)$ counterterms do not contribute.
This has two implications:

\noindent 1) the chiral meson loops are finite;

\noindent 2) these are the only contributions of $O(p^4)$ and
there is no dependence on unknown coupling constants.

The finiteness of the one--loop amplitude can be simply understood by the fact
that the superficial degree of divergence of the amplitude ($\sim \Lambda^2$,
with $\Lambda$ an ultra--violet cut--off)  is decreased by gauge invariance
together with the condition that the amplitude must vanish in the $SU(3)$
limit. Thus the pion loop amplitude is proportional
to $A(K_1^0 \rightarrow \pi^+\pi^-)$, i.e. to
$M_{K^0}^2 - M_{\pi^+}^2$,
while the kaon loop amplitude is proportional to $M_{K^0}^2 - M_{K^+}^2$
and thus can be neglected.

Adding the 8-- and 27--plet contributions one obtains \cite{DEG,BDM}
\beq
A(K_1^0\rightarrow 2 {\gamma}_{\parallel}) = {4\alpha F
\over \pi M^{2}_{K} }
( G_8 +{2\over 3}G_{27} ) ( {M}^{2}_{K} - {M}^{2}_{\pi} )
\big[ (q_{1}{\epsilon}_{2})(q_{2} {\epsilon}_{1})
-({\epsilon}_{1}{\epsilon}_{2})({q}_{1} {q}_{2})\big]H(0)~,
\label{eq:ksgga}
\eeq
where $q_1,\epsilon_1,q_2,\epsilon_2$ are the photon momenta and polarizations.
The function $H$,  defined in Appendix A, has an imaginary part
since the two pions can be on--shell.

The rate is given by
\beq
\Gamma({K}_{S}\rightarrow 2 {\gamma}_{\parallel}) =
{ {(G_8 +{2\over 3} G_{27})^2 ( {M}^{2}_{K} - {M}^{2}_{\pi} )^2\alpha^2 F^2}
\over{4 \pi^3 M_K } } \cdot
\left| H(0) \right|^2,
\eeq
which implies $\Gamma(K_S \rightarrow \gamma \gamma)=
1.5\cdot{10}^{-11} eV$. The branching ratio is
\beq
 BR(K_S \rightarrow \gamma \gamma)=2.1\cdot{10}^{-6},
\label{eq:ksggbr}
\eeq
corresponding to
\beq
{\Gamma({K}_{S}\rightarrow 2 \gamma)_{theor} \over
\Gamma({K}_{L}\rightarrow 2 {\gamma})_{exp}} \approx 2~.
\eeq
The experimental branching ratio \cite{NAKS}
\beq
BR({K}_{S}\rightarrow 2 \gamma)=(2.4 \pm 1.2)\cdot{10}^{-6}
\label{eq:ksggb}
\eeq
is in good agreement with the prediction (\ref{eq:ksggbr}).
This is a significant test for CHPT, since Eq.~(\ref{eq:ksggbr})
is unambiguously predicted to $O(p^4)$ in
terms of the $O(p^2)$ couplings $G_8$ and $G_{27}$.
Indeed, the absence of $O(p^4)$ counterterms not only implies
that the amplitude is finite but it also ensures that contact terms
will appear only
at $O(p^6)$ and are thus suppressed by the chiral expansion parameter
$\displaystyle{M_{K}^2 / (4\pi F_\pi)^2}\sim 0.2$.
This last statement is even more justified in this channel since there are no
vector meson exchange contributions at this order.
The effect of $\pi-\pi $ rescattering has been shown
to be negligible \cite{KaHo,Truong}.
Since the model--independent absorptive part is
dominant in this decay, one obtains similar results using a
phenomenological coupling in the pion loop  \cite{CCa,Uy}.

\subsection{ $K_L \rightarrow \pi ^0 \gamma \gamma$}
The general amplitude for $K_L \rightarrow \pi^0 \gamma \gamma$
is given by
\beq
M(K_L(p)\rightarrow \pi^0(p')\gamma (q_1,\varepsilon_1)
 \gamma (q_2,\varepsilon_2))=
\varepsilon_{1\mu}\varepsilon_{2\nu}M^{\mu\nu}(p,q_1,q_2)~,
\label{eq:klpgga}
\eeq
where $\varepsilon_1$, $\varepsilon_2$  are the photon polarizations.
If CP is conserved, $M^{\mu \nu}$ can be decomposed in two
invariant amplitudes:
\beqa
M^{\mu\nu}&=&{A(y,z) \over M_K^2}(q_2^{\mu}q_1^{\nu}-q_1\cdot q_2g^{\mu\nu})
\nonumber \\
& &+{2B(y,z) \over M_K^4}
(-p\cdot q_1 p\cdot q_2g^{\mu\nu}-q_1\cdot q_2 p^\mu p^\nu +
p\cdot q_1 q_2^\mu p^\nu + p\cdot q_2 p^\mu q_1^\nu)~,
\label{eq:klpggb}
\eeqa
with
\beq
y=p\cdot (q_1-q_2)/M_K^2~,~~~~~~~~~z=(q_1+q_2)^2/M_K^2~.
\label{eq:yz}
\eeq
Due to Bose symmetry $A(y,z)$ and $B(y,z)$ must be
symmetric for $q_1 \leftrightarrow q_2$ and consequently
depend only on $y^2$.

The physical region in the dimensionless variables y and z
is given by the inequalities
\beq
|y|\le {1\over2} \lambda^{1/2}(1,z,r_\pi^2)~,\qquad\qquad
0\le z \le (1-r_\pi)^2 ~,
\label{eq:phsp}
\eeq
where $r_\pi = M_\pi / M_K$ and the function $\lambda$ is
defined in Appendix A. From (\ref{eq:klpgga}) and (\ref{eq:klpggb})
we obtain the double differential decay rate for unpolarized photons:
\beq
 {d^2 \Gamma \over dy \ dz} = {M_K \over 2^9\pi^3} \left
\{z^2 \vert A+B \vert ^2 +
\left[y^2-{1\over 4}\lambda(1,z,r_\pi^2) \right]^2
\vert B \vert ^2 \right\}.
\label{eq:klpggc}
\eeq
We remark that, due to the different tensor structure in (\ref{eq:klpggb}),
the $A$ and
$B$ parts of the amplitude give rise to contributions to the differential decay
rate which have different dependence on the two--photon invariant mass $z$.
In particular, the second term in (\ref{eq:klpggc})  gives a non--vanishing
contribution to $\displaystyle{{d \Gamma(K_L \rightarrow \pi^0 \gamma \gamma)
\over dz}}$ in the limit $z \rightarrow 0$. Thus the kinematical region with
collinear photons is important to extract the $B$ amplitude.\par
\begin{figure}
    \begin{center}
       \setlength{\unitlength}{1truecm}
       \begin{picture}(10.0,7.0)
       \epsfxsize 10. true  cm
       \epsfysize 7.  true cm
       \epsffile{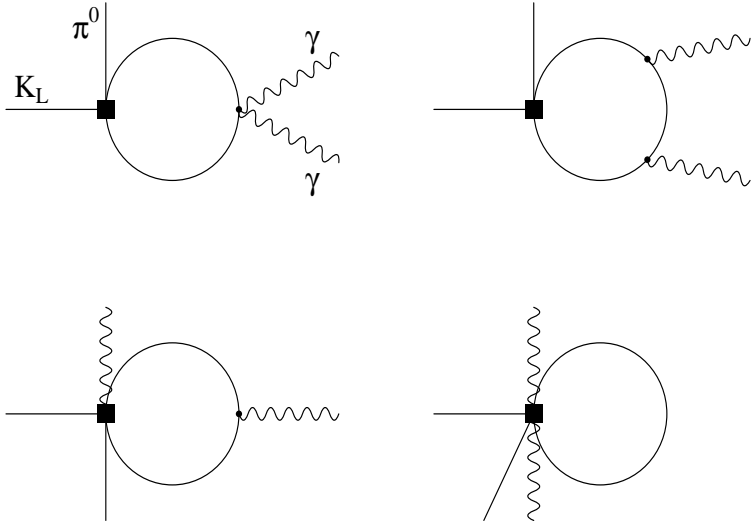}
       \end{picture}
    \end{center}
    \caption{One--loop diagrams for $K_L \to \pi^0 \gamma\gamma$
              in the diagonal basis of Refs.~\protect{\cite{EPRb,EPRc}}.}
    \protect\label{fig:klloops}
\end{figure}

We now consider this decay in the framework of CHPT.
Since $K_L$ and $\pi^0$ are neutral there is no tree--level $O(p^2)$
contribution. At $O(p^4)$, there are in principle
both loops and counterterms.
Since again $K_L$ and $\pi^0$ are neutral the latter ones do not
contribute, implying a finite one--loop amplitude \cite{EPRb,CD}.
The relevant diagrams
are shown in Fig.~\ref{fig:klloops}. At $O(p^4)$, the amplitude
$B$ vanishes because there are not enough powers of momenta.
The result for the amplitude $A$ (which at this order depends only on z)
due to the octet piece of the non--leptonic weak Lagrangian
is  \cite{EPRb,CD}:
\beq
A^{(8)}(z) = {{G_8\alpha M_K^2} \over \pi }
\left[(1-{r_\pi^2\over z})\cdot F\left({z \over r_\pi^2}\right)-
(1-{{r_\pi^2}\over z}-{1 \over z})F(z) \right] ~.
\label{eq:klpggd}
\eeq
The function $F(z)$, defined in Appendix A,
is real for $z \le 4$ and complex for $z \ge 4$. In (\ref{eq:klpggd}),
the contribution proportional to $F(z)$, which does not have
an absorptive part, comes from the kaon loop,  while the one
proportional to
$F({z \over r_\pi^2})$,  generated by
the pion loop, has an absorptive part since the pions can be on--shell.
Consequently, the kaon loop contribution is much smaller than the
pion one. For completeness we mention the contribution of the
$(27_L, 1_R)$ operator.
Due to the vanishing of the corresponding counterterms also this contribution
is finite and unambiguously predicted.
For the pion loop, which gives the larger contribution,
one obtains  \cite{CDM}:

\beq
A_{1/2}^{(27)}(z)={G_{27}\alpha M_K^2 \over 9\pi }
\left(1-{r_{\pi}^2\over z}\right) F\left({z \over r_\pi^2}\right)
\eeq
\beq
A_{3/2}^{(27)}(z)=-{5G_{27}\alpha M_K^2 \over 18 \pi }
\left[{3 - r_\pi^2 -14r_\pi^4-
(5-14r_\pi^2)z \over (1 -r_\pi^2)z}\right]
F\left({z \over r_\pi^2}\right) ~.
\eeq
Compared to the octet, there is a slight modification  of the spectrum and
of the width.

\begin{figure}
    \begin{center}
       \setlength{\unitlength}{1truecm}
       \setlength{\unitlength}{1truecm}
       \begin{picture}(8.0,8.0)
       \epsfxsize 16. true  cm
       \epsfysize 20.  true cm
       \epsffile{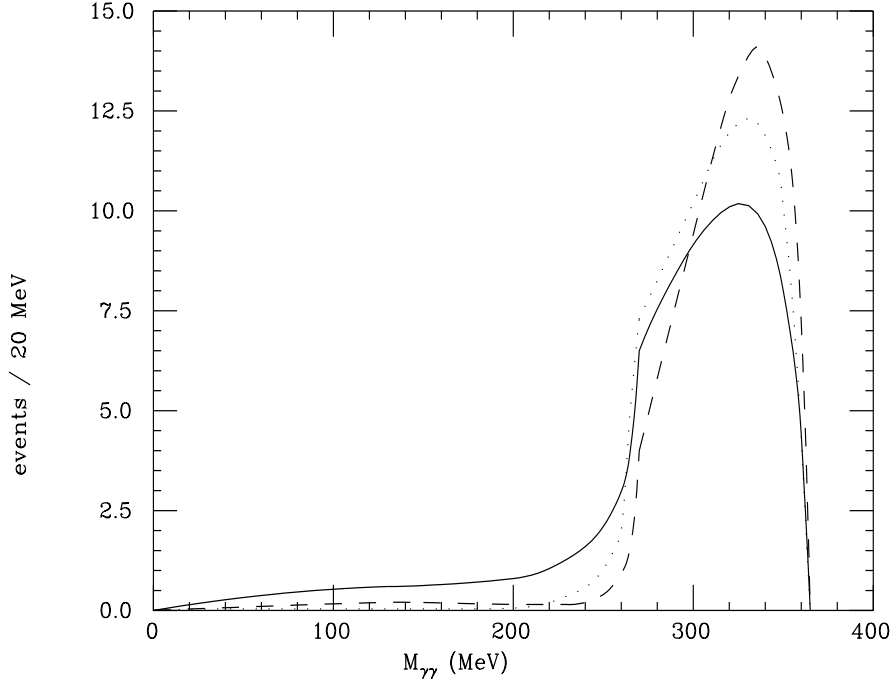}
       \end{picture}    \end{center}
    \caption{Theoretical predictions for the $2\gamma$ invariant mass
distribution in $K_L \rightarrow \pi^0 \gamma\gamma$.
The dotted curve is the $O(p^4)$ contribution,
the dashed and full curves correspond
to the $O(p^6)$ calculation of \protect{\cite{CEP}}
with $a_V=0$ and $a_V=-0.9$, respectively. The spectra are normalized to
the $50$ unambiguous events of NA31 (cf. Fig.
\protect{\ref{fig:klspexp}}).}
    \protect\label{fig:klspth}
\end{figure}

The z spectrum for a $y$--independent amplitude $A$ is given by:
\beq
{ d\Gamma \over dz} = {M_K \over 2^{10} \pi^3 } z^2
\lambda^{1/2}(1,z,r_\pi^2)\vert A(z)\vert ^2 ~.
\eeq
The $O(p^4)$ CHPT prediction for $\displaystyle{
{d\Gamma\over dz}(K_L \rightarrow \pi^0  \gamma\gamma)}$
is shown in Fig.~\ref{fig:klspth},
while the prediction for the branching ratio is:
\beq
BR^{(8)}(K_L \rightarrow \pi ^0 \gamma \gamma)=6.8\cdot{10}^{-7},\quad\qquad
BR^{(8+27)}(K_L \rightarrow \pi ^0 \gamma \gamma)=6.0\cdot{10}^{-7} ~.
\label{klpggbr}
\eeq

The spectrum predicted by CHPT to $O(p^4)$ is very characteristic:
a  peak in the absorptive region ($K_L\to \pi \pi\pi\to \pi \gamma \gamma$)
 and a negligible contribution at low $z$.
As we shall see,
contrary to the  $K_S \to \gamma\gamma$ case,
corrections to the $O(p^4$) CHPT prediction
for $K_L \to \pi^0\gamma\gamma$ can be sizeable \cite{EPRd}.
Phenomenological models  with large vector or scalar
exchange \cite{KOa,MI,Sb,HS92kl,Volkov}
can generate a larger branching ratio. However, due to the
presence of a $B$ amplitude, a non--vanishing contribution to the
spectrum at low $z$ is in general expected in such models.

Another point of interest of $K_L\rightarrow\pi^0 \gamma \gamma$
is its role as an intermediate state in $K_L\rightarrow\pi^0 e^+ e^-$.
The decay $K_L\rightarrow\pi^0 e^+ e^-$ has three kinds of contributions
 \cite{reviews,DHV,EPRd,DibDG,FR}: direct CP violation, indirect CP violation
via $K^0 - \bar{K^0}$ mixing (cf. $K_S\to \pi^0 e^+ e^-$ in Sect. 4)
and the CP conserving transition
$K_L\rightarrow\pi^0 \gamma^* \gamma^* \rightarrow\pi^0 e^+ e^- $.
The direct CP violating contribution is expected to give
$BR(K_L\rightarrow\pi^0 e^+ e^-) \simeq 10^{-11}$ \cite{Buras}.
At $O(p^4)$, the contribution of $K_L\rightarrow\pi^0 \gamma \gamma$
to this process is helicity suppressed via the amplitude A:
$BR(K_L\rightarrow\pi^0 e^+ e^-) \simeq 10^{-14}$.
However, at $O(p^6)$ the amplitude $B$ is generated.
Although this amplitude is a
higher--order effect in CHPT, the corresponding amplitude for
$K_L\rightarrow\pi^0 e^+ e^-$ is not helicity suppressed \cite{EPRc}
and could be substantially larger than the leading--order contribution
due to $A$. We stress again that the presence
of a $B$ amplitude could be checked experimentally by studying
the spectrum of $K_L\rightarrow\pi^0 \gamma \gamma$ at low $z$.

\begin{figure}
    \begin{center}
       \setlength{\unitlength}{1truecm}
       \begin{picture}(8.0,8.0)
       \epsfxsize 16. true  cm
       \epsfysize 20. true cm
       \epsffile{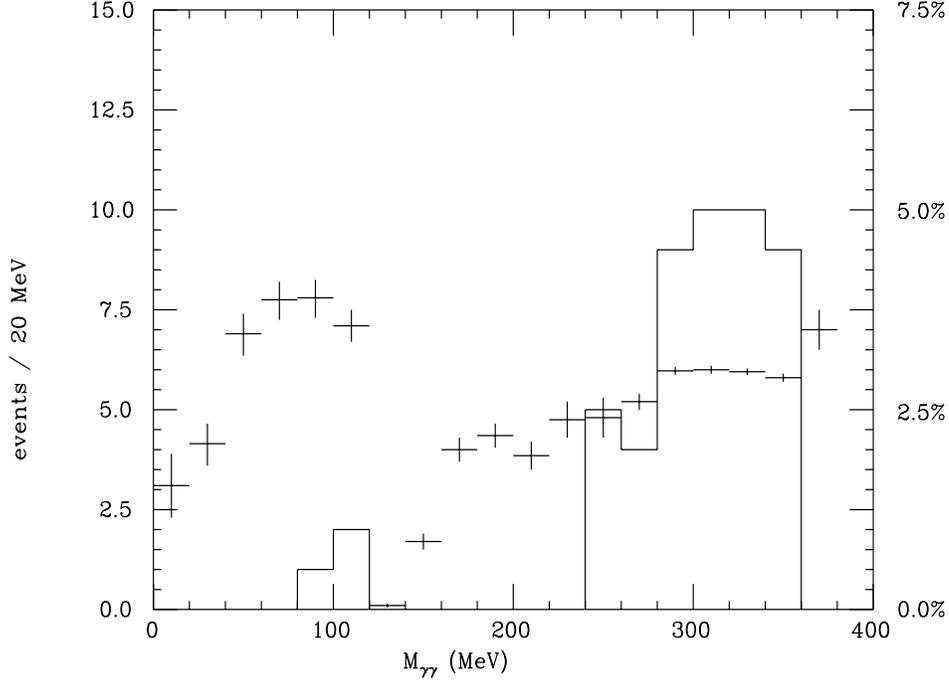}
       \end{picture}
    \end{center}
    \caption{$2\gamma$ invariant mass distribution for
unambiguous $K_L \rightarrow \pi^0 \gamma\gamma$ candidates
\protect{\cite{NAKL}} (solid histogram).
The crosses indicate the acceptance. }
    \protect\label{fig:klspexp}
\end{figure}

 From the experimental point of view,
an important background for this process is $K_L\rightarrow \pi^0 \pi^0$,
which makes it difficult to explore the region $z\sim {M_\pi^2 \over
M_K^2}$. The experimental situation is the following:
\beqa
BR(K_L \rightarrow \pi^0 \gamma \gamma)=& (1.7 \pm 0.2 \pm 0.2
)\cdot{10}^{-6}&\quad {\rm for}\quad M_{\gamma\gamma}>280~{\rm MeV}\quad {\rm
NA}31~ \cite{NAKL}, \label{eq:nakl}\\
BR(K_L \rightarrow \pi^0 \gamma \gamma)=&
(1.86\pm0.60\pm0.60)\cdot {10}^{-6}&
 \quad {\rm for}\quad M_{\gamma\gamma}>280~{\rm MeV}\quad {\rm
E}731~~\cite{FNALKL}, \label{eq:fnalkl}
 \eeqa
where $M_{\gamma\gamma}$ is the two--photon invariant mass:
\beq
 M_{\gamma\gamma}=\sqrt{(q_1+q_2)^2} ~.
\eeq
In Fig.~\ref{fig:klspexp} the NA31 two--photon invariant mass histogram is
displayed. We see that, although the rate seems underestimated, the number of
events for $M_{\gamma\gamma}<240$ MeV is very small, as predicted by
$O(p^4)$  CHPT. There is no evidence for a sizeable $B$ term.

Several contributions have been analysed that go beyond  O($p^4$).

The first and more controversial contributions are due to vector
(and/or  axial--vector and/or scalar) mesons. These can show up either
in the weak couplings
or in the  strong ones. While model predictions for the strong couplings
can be checked from experiments, this is not so easy for the weak ones.
Some phenomenological models can generate large vector \cite{MI,Sb,HS92kl} or
scalar exchange \cite{Volkov} contributions. One can
parametrize the local $O(p^6)$ contributions to
$K_L\rightarrow \pi^0 \gamma \gamma$, including direct weak transitions, by an
effective vector coupling $a_V$ \cite{EPRd} :
\beq
A={G_8 M_K^2\alpha\over \pi } a_V(3-z+r_{\pi}^2) ~,\qquad\qquad
B=-{2G_8 M_K^2\alpha\over \pi } a_V \label{klppge}~.
\eeq
Thus, $V$ exchange generates a $B$ amplitude changing the $O(p^4)$
spectrum, particularly in the region
of  small $z$, and contributing to the CP conserving part of $K_L\to\pi^0
e^+e^-$.
The $B$ amplitude gives rise to
\beq
BR(K_L \rightarrow \pi^0 e^+e^-)|_{abs}= 4.4\cdot {10}^{-12}a_V^2~.
\eeq
  From the analysis of events with $M_{\gamma\gamma}<240$
MeV, NA31 \cite{NAKL} have obtained the following limits:
\beq
-0.32 < a_V < 0.19  \qquad (90\%\ {\rm CL})\label{bound}.
\eeq

However, other $O(p^6)$ contributions have to be included \cite{Donoghue}.
Since  $G_8$ appearing in (\ref{eq:klpggd}) has been  extracted
from $K_S\to 2\pi$  underestimating
the $K_L\to 3\pi$ amplitudes  by 20$\%$--$30\%$, the authors of
\cite{CDM,CEP} included the O($p^4)$ corrections to $K_L\to 3\pi$ for
the weak vertex in Fig.~\ref{fig:klloops}.
The main feature of the analysis of \cite{CDM,CEP} is an increase of the width
by some $25\%$ and a modification of the spectrum. In fact, the resulting
spectrum is even more strongly peaked at large $z$ than found
experimentally (see Fig.~\ref{fig:klspth}).
However, the contribution  of vector mesons in (\ref{klppge})
still has to be added. Choosing $a_V$ to reproduce the experimental
spectrum leads to an additional increase of the rate \cite{CEP} in
good agreement with the measured values (\ref{eq:nakl}), (\ref{eq:fnalkl}).
The corresponding contribution to $K_L \rightarrow \pi^0 e^+e^-$
($2\gamma$ absorptive part only) is still
smaller than the direct CP violating contribution \cite{CEP}. Finally,
a more complete unitarization of the $\pi-\pi$ intermediate states
(Khuri--Treiman treatment) and the inclusion of the experimental
$\gamma\gamma\to \pi^0\pi^0$ amplitude increases the
$K_L\to \pi^0\gamma\gamma$ width by another 10\% \cite{KaHo}.

DA$\Phi$NE seems an ideal machine to investigate the relative role
of CHPT and VMD in $K_L\rightarrow \pi^0\gamma\gamma$,
establishing  both the absorptive and the
dispersive parts of the decay amplitude. In particular, the region of collinear
photons could be definitely assessed.

\subsection{ $K_L \rightarrow \gamma \gamma$}

\begin{figure}
    \begin{center}
      \setlength{\unitlength}{1truecm}
       \begin{picture}(5.0,3.0)
       \epsfxsize 5. true  cm
       \epsfysize 3. true cm
       \epsffile{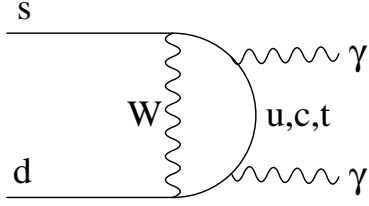}
       \end{picture}
    \end{center}
    \caption{ Short--distance contribution to $K_L \to \gamma\gamma$.}
    \protect\label{fig:klggshort}
\end{figure}

\begin{figure}
    \begin{center}
       \setlength{\unitlength}{1truecm}
       \begin{picture}(5.0,3.0)
       \epsfxsize 5.0  true  cm
       \epsfysize 3.  true cm
       \epsffile{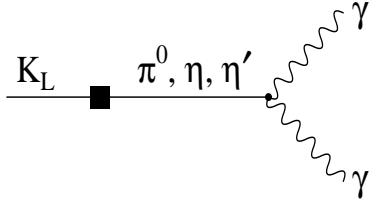}
       \end{picture}
    \end{center}
    \caption{Long--distance contribution to $K_L \to \gamma\gamma$.}
    \protect\label{fig:klgganom}
\end{figure}

If CP is conserved, $K_L$ decays into two photons with perpendicular
polarizations $(2\gamma_\perp)$. DA$\Phi$NE could improve the measurement
of the decay width. However, since the theory for this channel
is afflicted by several uncertainties, it will be difficult
 to achieve  new insight from this measurement alone.
This decay has been  historically very
important for understanding the GIM mechanism  \cite{GW80}.
The interplay between the
short--distance contributions in Fig.~\ref{fig:klggshort}
and the long--distance contribution
in Fig.~\ref{fig:klgganom}
is a matter of past and current interest  \cite{MP,DHL,Uy}. This
is also related to the study of CP
violation in this channel at LEAR  \cite{Martin,CCa,BDM,Uy,EP}.
The experimental width $\Gamma (K_L \rightarrow
\gamma\gamma)$ is needed to predict direct CP violation in this
channel. More generally, it will be useful for the theoretical
analysis of other decays (cf. Sect.~\ref{sec:Dalitz}).

The loop integral of the short--distance contribution in
Fig.~\ref{fig:klggshort} is a function of
$m_i^2 / M_W^2$, where $m_i$ is the mass of the intermediate quark.
 The contributions for $m_i=0$ cancel when we sum over
all $u-$like quarks (GIM mechanism).
Also for $m_i \neq 0$ the short--distance contributions are negligible
compared to the long--distance ones  \cite{GW80,MP}. Thus,
the main contribution is expected to come from long distances. In
the framework of CHPT, the Wess--Zumino term and the
$\Delta S=1$ weak Lagrangian generate the CP conserving amplitude.
At $O(p^4)$ in CHPT one has:
\beqa
A(K_L \rightarrow 2\gamma_{\perp})_{O(p^4)} &=&
A(K_L \rightarrow \pi^0 \rightarrow 2\gamma_{\perp}) +
A(K_L \rightarrow \eta_8 \rightarrow 2\gamma_{\perp}) \nonumber\\
&=& A(K_L \rightarrow \pi^0) A(\pi^0 \rightarrow 2\gamma_{\perp})
\left[{1 \over M_K^2 - M_\pi^2} + {1 \over 3}\cdot{1 \over M_K^2 - M_8^2}
\right]
 \simeq 0 ~.\label{Klp4}
\eeqa


The Gell-Mann--Okubo formula, which holds at this order, tells us that the
amplitude (\ref{Klp4}) is zero at $O(p^4)$. If we write the full
amplitude as
\begin{equation}
A(K_L \rightarrow 2\gamma_{\perp})=
A(K_L \rightarrow \pi^0) A(\pi^0 \rightarrow 2\gamma_{\perp})\dfrac{F_2}
{M_K^2}\label{ftwo}~,
\end{equation}
one would naively expect $F_2\sim M_K^2/(4\pi F)^2\simeq 0.2$. However,
comparison with experiment shows that ${F_2}^{(exp)}\simeq 0.9$ implying
large higher--order corrections in CHPT. At $O(p^6)$, the
$\eta'$ pole appears and contributes to the large value
of  ${F_2}^{(exp)}$. The pole contributions to the amplitude can
be written as \cite{GW80,MP,DHL,DEG,CCa,Chengb,BDM}
\beq
A(K^{0} \rightarrow 2\gamma_{\perp})=\sum_{P={\pi}^{0},\eta ,
{\eta}^{\prime}}{ A(K^0 \rightarrow P) \over{{M}_{K}^{2}-{M}_{P}^{2}}}
A(P \rightarrow 2\gamma_{\perp}) ~.
\eeq
Including the $\eta-\eta'$
mixing angle $\Theta$, $F_2 $ defined in (\ref{ftwo}) is given by
\beq
F_2 = \dfrac{1}{1 - r_\pi^2} -
\dfrac{(c_\Theta - 2\sqrt{2}\,s_\Theta)(c_\Theta + 2\sqrt{2}\,
\rho s_\Theta)}{3(r_\eta^2 - 1)} +
\dfrac{(2\sqrt{2}\,c_\Theta + s_\Theta)(2\sqrt{2}\,
\rho c_\Theta -s_\Theta)}{3(r_{\eta'}^2 -1)}\label{ftwoexp}
\eeq
$$
r_i = M_i/M_K, \qquad c_\Theta = \cos \Theta, \qquad s_\Theta = \sin \Theta.
$$
  $\rho \neq 1$ takes into account possible deviations
from nonet symmetry for the non--leptonic weak vertices (nonet symmetry
is assumed for the strong WZW vertices).
The $1/N_c$ prediction  $\Theta \simeq - 20^\circ$ \cite{GaLe} and nonet
symmetry agree very well with the $2\gamma$ decay widths of $\pi^0,\eta,\eta'$
\cite{Kawa}.

The $\eta$ and $\eta'$ contributions in (\ref{ftwoexp}) interfere destructively
for $0 \leq \rho < 1$ and $\Theta \simeq - 20^\circ$. $F_2$ is
dominated by the pion pole (cf. the similar situation for $K_L\to
\pi^+\pi^-\gamma$ discussed in Sect.~\ref{sec:1g}).
Since there are certainly other contributions
of $O(p^6)$ and higher, Eq.~(\ref{ftwoexp}) at best contains the dominant
contributions.

\subsection{ $K^+ \rightarrow \pi^+ \gamma \gamma $}
At present there is only an upper bound for the branching ratio
of this process \cite{Atiya},
which depends upon the shape of the spectrum due to the different experimental
acceptance:

\beq
BR(K^+ \rightarrow \pi^+ \gamma \gamma)_{exp}
 \le 1.5 \cdot 10^{-4}\quad {\rm CHPT \ amplitude}.
\label{limkpgg}\eeq
\beq
BR(K^+ \rightarrow \pi^+ \gamma \gamma)_{exp}
 \le 1.0 \cdot 10^{-6}\quad {\rm constant \ amplitude}.
\eeq

A cut in the two--photon invariant mass is necessary
to disentangle this channel from the background $K^+ \rightarrow \pi^+ \pi^0
\rightarrow \pi^+ \gamma \gamma $.\par
Gauge invariance and chiral symmetry imply that this decay starts at
$O(p^4)$ in CHPT [cf. Eq.~(\ref{eq:pig*})].
Two invariant amplitudes contribute at this order:
\beqa
&&M(K^+(p) \rightarrow \pi^+(p') \gamma(q_1,\epsilon_1) \gamma(q_2,\epsilon_2)
 ) = \nonumber \\ \qquad
&&\qquad=\epsilon_\mu(q_1) \epsilon_\nu(q_2)
\left[ A(y,z)
\dis{ (q_2^{\mu}q_1^{\nu}-q_1 \cdot q_2g^{\mu\nu}) \over  M_K^2} +
 C(y,z)\varepsilon^{\mu\nu\alpha\beta}
\dis{q_{1\alpha}q_{2\beta}  \over M_K^2}
\right],
\eeqa
where $y$, $z$ have been defined in (\ref{eq:yz}) and
the physical region for $y$, $z$ is given in (\ref{eq:phsp}).
The amplitude $A$ corresponds to a $2\gamma$ state with $CP = +1$,
while $C$ corresponds to a state with $CP = -1$.
Compared to (\ref{eq:klpggb}), the new amplitude $C(y,z)$ appears
because initial and final states are not CP eigenstates.
Since $A$ and $C$ depend only on z at $O(p^4)$, we  integrate  over $y$
to obtain the following expression for the
two--photon invariant mass spectrum:
\beq
 {d\Gamma \over dz}(K^+ \rightarrow \pi^+ \gamma \gamma) =
 {M_K \over 2^{10} \pi^3} z^2
\lambda^{1 \over 2}(1,z,r_\pi^2)
(|A(z)|^2 + |C(z)|^2)~.
\label{eq:kppggd}
\eeq

\begin{figure}
    \begin{center}
       \setlength{\unitlength}{1truecm}
       \begin{picture}(4.5,4.0)
       \epsfxsize 4.5 true  cm
       \epsfysize 4.  true cm
       \epsffile{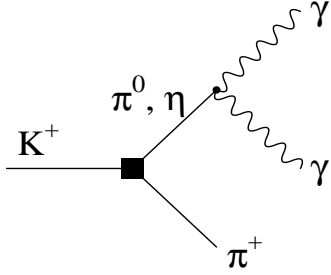}
       \end{picture}
    \end{center}
    \caption{Pole diagram for $K^+ \rightarrow \pi^+ \gamma
\gamma$.}
    \protect\label{fig:anomkppgg}
\end{figure}

The calculation for this decay proceeds similarly to the case of
$K_L \rightarrow \pi^0 \gamma \gamma$. The crucial
difference is that now the $O(p^4$) counterterms do not vanish
since here the external kaon and pion are charged. Loops and counterterms
contribute to $A$, while the reducible anomalous amplitude
(cf. Sect.~\ref{sec:intro}) of Fig.~\ref{fig:anomkppgg} contributes to $C$.

The loop contribution  turns out to be finite so that
the total counterterm contribution must be scale independent.
One finds  \cite{EPRc}
\beqa
A(z) &=& {G_8 M_K^2 \alpha \over 2 \pi z}\left[(r_\pi^2-1-z)
F\left({z \over r_\pi^2}\right) +(1 - r_\pi^2 -z)F\left(z\right)
+ \hat{c}z\right], \label{eq:kppgga}\\
C(z) &=& {G_8 M_K^2 \alpha \over \pi }
\left[{z-r_\pi^2 \over z - r_\pi^2
+ir_\pi{\Gamma_{\pi^0} \over M_K}} -  {z-{2+r_\pi^2 \over 3} \over
 z - r_\eta^2}\right], \label{eq:kppggb}
\eeqa
where $\hat{c}$ is an unknown coupling constant due to
$O(p^4)$ counterterms:
\beq
\hat{c} ={ {128 \pi^2}\over{3}}[3(L_9+L_{10}) +N_{14}-N_{15}-2N_{18}] ~.
\label{eq:c-hat}
\eeq
$L_9+L_{10}$ and $N_{14}-N_{15}-2N_{18}$ are separately scale independent.
In (\ref{eq:kppgga}), the term proportional to $F(z)$ comes from the kaon loop
and does not have an
absorptive part, while the term with $F({z \over r_\pi^2})$ is
due to the pion loop and has an absorptive part. As for $K_L\to \pi^0
\gamma\gamma$, the kaon loop contribution is much smaller than
the pion loop one.
\begin{figure}
    \begin{center}
       \setlength{\unitlength}{1truecm}
       \setlength{\unitlength}{1truecm}
       \begin{picture}(8.0,8.0)
       \epsfxsize 16. true  cm
       \epsfysize 20.  true cm
       \epsffile{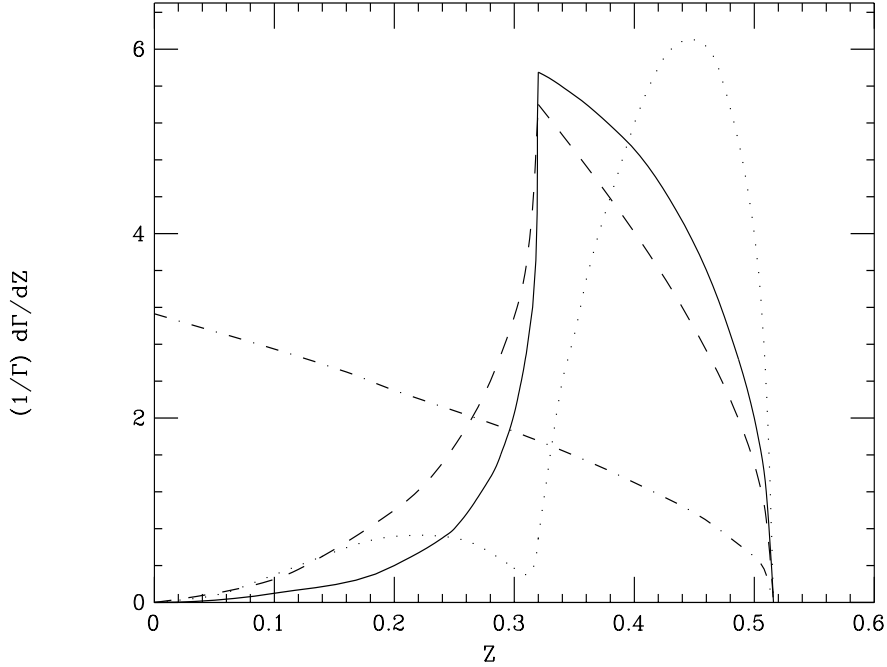}
       \end{picture}
    \end{center}
    \caption{Normalized theoretical $z-$distribution in $K^+ \rightarrow
\pi^+ \gamma \gamma$ \protect{\cite{EPRc}},
for several values of $\hat{c}$: $\hat{c}=0$ (full curve), $\hat{c}=4$
(dashed curve) and $\hat{c}=-4$ (dotted curve). The dash--dotted curve is
the phase space. }
    \protect\label{fig:kppggsp}
\end{figure}

Using (\ref{eq:kppgga}) and  (\ref{eq:kppggb}) in (\ref{eq:kppggd}),
one obtains
\beqa
&\Gamma_A(K^+ \rightarrow \pi^+\gamma\gamma) &= (2.80+0.87\hat{c}+
0.17\hat{c}^2)\cdot 10^{-20}\ {\rm MeV} \label{eq:kppggc}\\
&\Gamma_C(K^+ \rightarrow \pi^+\gamma\gamma) &= 0.26
\cdot 10^{-20}\ {\rm MeV}
\eeqa
showing the dominance of the loop over the anomalous contribution.
Since $\hat{c}$ is unknown  we can deduce from (\ref{eq:kppggc})
only a lower bound for the rate, which is obtained for $\hat{c}=-2.6$:
\beq
\Gamma(K^+ \rightarrow \pi^+\gamma\gamma)=\Gamma_A+\Gamma_C \ge 2 \cdot
10^{-20} \ {\rm MeV},
\eeq
or equivalently
\beq
BR(K^+ \rightarrow \pi^+\gamma\gamma) \ge 4 \cdot 10^{-7}~.
\label{eq:kppggcp}
\eeq
The spectrum (\ref{eq:kppggd}) is predicted up to the unknown
parameter $\hat{c}$.
Experiments can test the predicted shape and constrain the possible
values of $\hat{c}$. The shape is in fact
very sensitive to the value of $\hat{c}$ (see Fig.~\ref{fig:kppggsp}).
In the factorization model (cf. Ref.~\cite{EKW} and references therein)
one obtains
\beq
\hat{c} = {(4\pi F_\pi)^2\over M_\rho^2}(1 - 2 k_f) = 2.3 (1- 2 k_f)
\eeq
where $k_f= O(1)$ is a free parameter of the model. As a special
case, the weak deformation model \cite{EPRd} has $k_f=1/2$ and
$\hat{c}=0$ implying
$BR(K^+ \rightarrow \pi^+\gamma\gamma)=6\cdot10^{-7}$. Naive
factorization corresponds to $k_f=1$ and $\hat{c}=-2.3$ with
$BR(K^+ \rightarrow \pi^+\gamma\gamma)=4\cdot10^{-7}$.

\subsection{  $K_S \rightarrow \pi^0 \gamma \gamma $}
This decay proceeds through a reducible anomalous amplitude
similar to the one in  Fig.~\ref{fig:anomkppgg},
where $K^+$ and $\pi^+$ are replaced  respectively by
$K_S$ and $\pi^0$ \cite{EPRb}.
We do not include $\eta-\eta'$ mixing, which is a higher--order effect in CHPT.
Actually, this process is mainly sensitive to the $\pi^0$ pole, probing
the momentum dependence of the $K^0 \pi^0 \pi^0$ vertex.
Due to the strong background coming from $K_S\rightarrow \pi^0 \pi^0$,
a  cut in $z$ has to be applied.

\begin{figure}
    \begin{center}
       \setlength{\unitlength}{1truecm}
       \setlength{\unitlength}{1truecm}
       \begin{picture}(8.0,8.0)
       \epsfxsize 16. true  cm
       \epsfysize 20.  true cm
       \epsffile{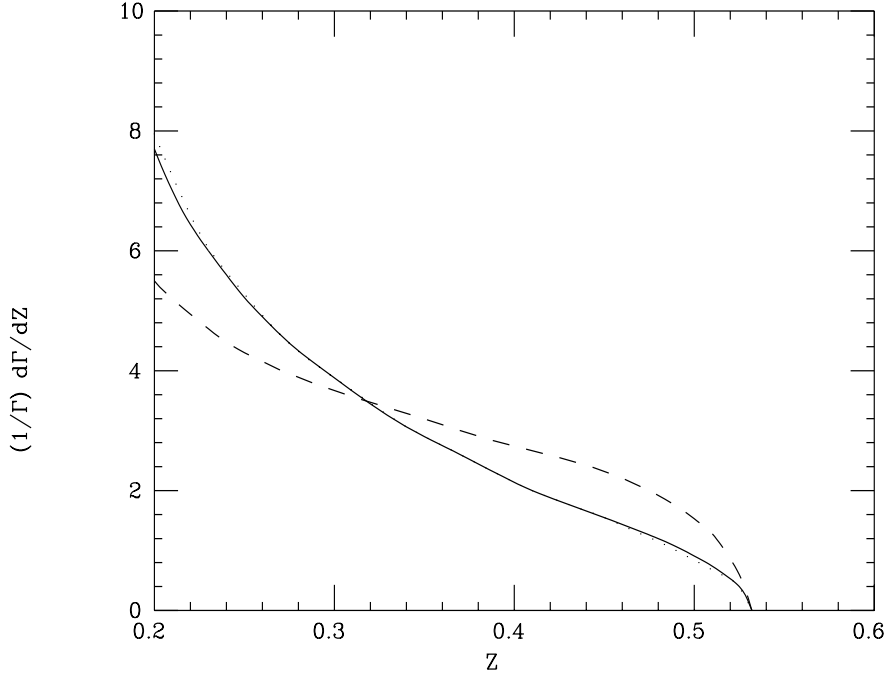}
       \end{picture}
    \end{center}
    \caption{Normalized theoretical $z-$distribution in $K_S \rightarrow
\pi^0 \gamma \gamma$ \protect{\cite{EPRb}}
in the region $0.2\leq z\leq (1-r_\pi)^2$ (full curve).
The pion pole contribution alone is given by the dotted curve. The dashed curve
is obtained assuming momentum independent weak vertices.}
    \protect\label{fig:kspggsp}
\end{figure}

The amplitude is given by \cite{EPRb}:
$$
 M(K_S(p) \rightarrow \pi^0(p') \gamma(q_1,\epsilon_1) \gamma(q_2,\epsilon_2))
=C(z) \varepsilon^{\mu\nu\alpha\beta} {q_{1\alpha}q_{2\beta}  \over M_K^2}
\epsilon_\mu(q_1) \epsilon_\nu(q_2)
$$
\beq
C(z)= {G_8 \alpha M_K^2\over \pi}
\left[{2-z-r_\pi^2 \over z - r_\pi^2 +ir_\pi \Gamma_{\pi^0}/ M_K}
-  {2-3z+r_\pi^2 \over 3 ( z - r_\eta^2 +ir_{\eta} \Gamma_{\eta}/
 M_K)}\right],
\eeq
where $r_i={M_i \over M_K}$. The $2\gamma$ invariant mass
spectrum is dominated by the $\pi^0$ pole and is given by
\beq
{d\Gamma \over dz}(K_S \rightarrow \pi^0 \gamma \gamma) =
 {M_K \over 2^{10} \pi^3} z^2
\lambda^{1 \over 2}(1,z,r_\pi^2)
|C(z)|^2~.
\eeq
This spectrum and the one obtained with a constant weak coupling
are shown in Fig.~\ref{fig:kspggsp}, where the cut $z\ge 0.2$ has been
applied. The branching ratio with this cut is
\beq
BR(K_S\rightarrow\pi^0 \gamma \gamma)_{z\ge 0.2} =
3.8\cdot 10^{-8}.
\eeq
DA$\Phi$NE should be able to see this decay, but it will be very difficult
to confirm the kinematical dependence of the vertex.

\subsection{  $K_{L,S} \rightarrow \pi^0\pi^0 \gamma \gamma $ }
These decays can be unambiguously predicted in CHPT at $O(p^4)$.
For both transitions there are no
$O(p^4)$ weak counterterms and thus the one--loop amplitude
is finite.  $A(K_L\rightarrow \pi^0\pi^0 \gamma \gamma)$
is a reducible anomalous amplitude \cite{DFR},
while $A(K_S \rightarrow \pi^0\pi^0 \gamma \gamma)$
is a pure loop amplitude \cite{FK93}.

 From the experimental point of view, it is convenient for both decays
to perform a cut in the two--photon invariant mass
$M_{\gamma\gamma}$ in order to avoid the background from
$K_{L,S} \rightarrow \pi^0\pi^0 \pi^0$,
$K_{L,S} \rightarrow \pi^0\pi^0$ and $K_{L,S} \rightarrow \pi^0\pi^0
\gamma $. Following Ref.~\cite{FK93}, we report in Table~\ref{tab:kls00}
the predictions for the two branching ratios, with cuts
$M_{\gamma\gamma} >\delta m$ and $|M_{\gamma\gamma}-M_\pi| >\delta m$,
for different values of $\delta m$. Since
$K_L\rightarrow \pi^0\pi^0 \gamma \gamma$ is dominated by the pion pole
its branching ratio depends strongly on the value of the cut $\delta m$.

\begin{table}
\label{tab:kls00}
\[ \begin{array}{|c|c|c|} \hline \delta m~ \rm{(MeV)} &
 BR(K_{S} \rightarrow \pi^0\pi^0 \gamma \gamma) &
 BR(K_{L} \rightarrow \pi^0\pi^0 \gamma \gamma)  \\ \hline
5 	& 5.6\cdot 10^{-9} & 2.0 \cdot 10^{-7} \\ \hline
10 	& 5.3\cdot 10^{-9} & 8.4 \cdot 10^{-8} \\ \hline
20 	& 4.7\cdot 10^{-9} & 3.0 \cdot 10^{-8} \\ \hline
30 	& 4.0\cdot 10^{-9} & 1.4 \cdot 10^{-8} \\ \hline
40 	& 3.4\cdot 10^{-9} & 6.4 \cdot 10^{-9} \\ \hline
\end{array}\]
\caption{$K_{L,S} \rightarrow \pi^0\pi^0 \gamma \gamma$ branching ratios for
 different values of the cuts in the two--photon invariant mass
($M_{\gamma\gamma} >\delta m$ and $|M_{\gamma\gamma}-M_\pi| >\delta m$).}
\end{table}

DA$\Phi$NE should be able to see both these decays, which have not
been observed yet.

\subsection{  $K^+ \rightarrow \pi^+\pi^0 \gamma \gamma $ }

We finally mention this decay which will probably not be observed at
DA$\Phi$NE. Differently from the two--photon decays discussed above,
a bremsstrahlung amplitude is present in this case. In agreement with
the general statement of Eq.~(\ref{brems}),
this is the only contribution in CHPT at $O(p^2)$.

 From the experimental point of view, it is again necessary to make a
cut in the two--photon invariant mass in order to avoid the background from
$K^+\to\pi^+\pi^0\pi^0$. For $M_{\gamma\gamma}>170$ MeV the
bremsstrahlung amplitude, which is suppressed by the
$\Delta I=1/2$ rule, leads to a  branching ratio of the order of  $10^{-10}$
\cite{ENPb}.

At $O(p^4)$ in CHPT, there is also a direct emission amplitude
which, analogously to the  $K^+ \rightarrow \pi^+\gamma \gamma $ case,
has both anomalous and non--anomalous contributions. The latter depend
on the scale independent combinations
\beq
N_{14}-N_{15}-N_{16}-N_{17}\quad {\rm and} \quad N_{14}-N_{15}-2N_{18}~,
\label{ctc}
\eeq
while the former depend on the combination $a_3-a_2/2$
\cite{ENPb} (see Sect. \ref{sec:intro}).
Although it would be very useful to extract the
counterterm combinations (\ref{ctc}) from this decay,
the direct emission and the interference amplitudes
lead to branching ratios  which are  of $O(10^{-10})$
\cite{ENPb}. DA$\Phi$NE will probably only put an upper limit
on this decay.

\subsection{CP violation }

The charge asymmetry in the decays $K^\pm \to \pi^\pm \gamma\gamma$
is an interesting observable for detecting direct CP violation. The amplitude
for $K^- \to \pi^- \gamma\gamma$ is obtained from (\ref{eq:kppgga})
and (\ref{eq:kppggb}) by replacing $G_8$ and $\hat{c}$ (defined in
(\ref{eq:c-hat}))
by their complex conjugates. The interference between the imaginary part
of $\hat{c}$ and the CP invariant absorptive part of the amplitude
(\ref{eq:kppgga}) generates a charge asymmetry \cite{EPRc}
\beqa
\Delta \Gamma &=&\Gamma(K^+ \rightarrow \pi^+\gamma\gamma)-
  \Gamma(K^- \rightarrow \pi^-\gamma\gamma) \nonumber \\
&=& \dis{\Imm\hat{c}|G_8 \alpha|^2 M_{K^+}^5 \over 2^{10} \pi^5 }
\int_{4 r_\pi^2}^{(1-r_\pi)^2} dz \lambda^{1 \over 2}(1,z,r_\pi^2)
(r_\pi^2-1-z)z  \Imm F (z / r_\pi^2) \nonumber \\
&\simeq& 1.5\cdot 10^{-20} \Imm \hat{c}\ ~{\rm MeV}~.
\eeqa
For an estimate of $\Delta \Gamma$ we need some information about the
imaginary part of the counterterm combination $\hat{c}$. It is clear that
$L_9$ and $L_{10}$ cannot contribute to this quantity. Furthermore, the
long--distance contribution to the phase of the weak counterterm is already
included in $G_8$. Therefore, $\Imm\hat{c}$ is governed by the short--distance
contribution to  $N_{14}-N_{15}-2N_{18}$  \cite{EPRc}, mainly
due to the electromagnetic penguin operator \cite{GW80}. The
original estimate of Ref.~\cite{EPRc} included only the
contribution to $\Imm N_{14}$ leading to the order--of--magnitude
estimate
\beq
|\Imm \hat{c}| \sim 3 \cdot 10^{-3}~.
\label{eq:optimc}
\eeq
However, it was later shown by Bruno and Prades \cite{BP93} that
there is also a contribution to $\Imm N_{18}$ which in fact cancels
$\Imm N_{14}$ to leading order in ${1 / N_c}$. Thus, the charge
asymmetry is probably much smaller than the original estimate of
Ref.~\cite{EPRc} and certainly beyond reach for DA$\Phi$NE:
\beq
{\Delta \Gamma (K^\pm \rightarrow \pi^\pm \gamma\gamma) \over
2 \Gamma(K^\pm \rightarrow \pi^\pm \gamma\gamma)} \ll 10^{-3}~.
\eeq

CP violation in $K_L\to \gamma\gamma$ could also
be interesting \cite{reviews,Golowich}.
As for $K \to \pi\pi$, one can separate  direct and indirect
CP violating contributions defining
\beq {\eta}_{\parallel}={{A(K_L \rightarrow 2\gamma_\parallel)}\over
{{A(K_{S} \rightarrow 2\gamma_{\parallel })}}}
\equiv \epsilon+ \epsilon_{\gamma \gamma \parallel}^\prime~,
\qquad
{\eta}_{\perp}={{A(K_{S} \rightarrow 2\gamma_{\perp })}\over
{{A(K_{L} \rightarrow 2\gamma_{\perp })}}}
\equiv \epsilon+ \epsilon_{\gamma \gamma \perp }^\prime~.
\eeq
While $\epsilon_{\gamma \gamma \parallel }^\prime \simeq
\epsilon^\prime_{\pi\pi}$, the direct CP violating parameter
$\epsilon_{\gamma \gamma \perp }^\prime$  could be substantially
larger than  $\epsilon^\prime_{\pi\pi}$ \cite{CCa,BDM,EP}.
As shown in  \cite{Fuka,Pugliese}, it is possible to
study CP violation through time asymmetries in this channel, but statistics
at DA$\Phi$NE does not seem to be large enough.

\subsection{ Improvements at DA$\Phi$NE }

The measurements of the $K_S\to \gamma\gamma$ and
$K_L\to\pi^0 \gamma\gamma$ branching ratios will be sensitive to
contributions of $O(p^6)$ in CHPT. In particular, the study of the
$K_L\to\pi^0 \gamma\gamma$
spectrum will be very useful for understanding the role of resonances, the
convergence of the CHPT expansion
and for improving the predictions for the CP conserving
contributions to  $K_L\to\pi^0 e^+e^-$.

In the decay $K^+ \to \pi^+ \gamma\gamma$, it will be important
to determine the $O(p^4)$ counterterm combination $N_{14}-N_{15}-2N_{18}$
and to check the correlation between the rate and the shape of
the $2\gamma$ spectrum.

Finally, the detection of the as yet unobserved decays
$K_S \to \pi^0 \gamma\gamma$ and $K_{L,S} \to \pi^0\pi^0 \gamma\gamma$
will furnish new important tests of CHPT, both in the anomalous and
in the non--anomalous sectors.

\setcounter{equation}{0}
\setcounter{subsection}{0}

\section{Kaon decays with one photon in the final state}
\label{sec:1g}

\subsection{Matrix elements and decay rates}
The amplitude for $K(P) \ra \pi_1(p_1) + \pi_2(p_2) + \gamma(q)$
is decomposed into an electric amplitude $E(x_i)$ and a magnetic
amplitude $M(x_i)$:
\beq
A(K \ra \pi_1 \pi_2 \gamma) = \ve^\mu(q)^* [E(x_i)(p_1 q \, p_{2\mu} -
p_2 q \, p_{1\mu}) + M(x_i) \ve_{\mu\nu\rho\sigma} p_1^\nu p_2^\rho
q^\sigma]/M_K^3~, \label{adec}
\eeq
\nopagebreak
$$
x_i = \dfrac{P p_i}{M_K^2} \qquad (i = 1,2), \qquad
x_3 = \dfrac{Pq}{M_K^2}, \qquad x_1 + x_2 + x_3 = 1.
$$
The invariant amplitudes $E(x_i)$, $M(x_i)$ are dimensionless. Summing
over the photon helicity, the differential decay distribution can be
written as ($r_i = M_{\pi_i}/M_K$)
\beqa
\dfrac{d\Gamma}{dx_1 dx_2} &=&
\dfrac{M_K}{4(4\pi)^3} (|E(x_i)|^2 + |M(x_i)|^2) \cdot \no \\
&& [(1 - 2x_3 - r_1^2 - r_2^2)(1 - 2x_1 + r_1^2 - r_2^2)
(1 - 2x_2 + r_2^2 - r_1^2) \no \\
&& - r_1^2(1 - 2x_1 + r_1^2 - r_2^2)^2- r_2^2(1 - 2x_2 + r_2^2 - r_1^2)^2]~.
\eeqa
There is no interference between $E$ and $M$ as long as the photon helicity
is not measured.

\subsection{Low's theorem and the chiral expansion}
\label{subsec:Low}
The behaviour of the electric amplitude in the limit of small photon energies
($E_\gamma \ra 0$) is governed by Low's theorem \cite{Low}. $E(x_i)$ can be
written as the sum of a bremsstrahlung part $E_B(x_i)$ and a direct emission
part $E_{DE}(x_i)$,
\beq
E(x_i) = E_B(x_i) + E_{DE}(x_i)~,
\eeq
where the dependence on the photon energy is of the form
\beq
E_B(x_i) \sim 1/E^2_\gamma, \qquad \qquad  E_{DE}(x_i) = {\rm const.}
+ O(E_\gamma)~.
\eeq
The bremsstrahlung term $E_B(x_i)$ is given by
\beqa
E_B(x_i) &=& \dfrac{e M_K^3 A(K \ra \pi_1 \pi_2)}{Pq} \left(\dfrac{Q_2}{p_2q}
- \dfrac{Q_1}{p_1q}\right) \no \\
&=& \dfrac{2 e A(K \ra \pi_1 \pi_2)}{M_K x_3} \left(\dfrac{Q_2}{1 + r_1^2
- r_2^2 - 2x_1} - \dfrac{Q_1}{1 + r_2^2 - r_1^2 - 2x_2}\right) ~,
\eeqa
where $e Q_i$ is the electromagnetic charge of $\pi_i$. $E_B(x_i)$ is thus
completely determined by the amplitude for the decay $K \ra \pi_1 \pi_2$.

This shows that to lowest $O(p^2)$ in CHPT the $K \ra \pi_1 \pi_2 \gamma$
amplitudes are completely determined by $E_B(x_i)$. In other words, there is
no additional information at $O(p^2)$ that would not already be contained in
the corresponding non--radiative transitions $K \ra \pi_1 \pi_2$.
Contributions to $E_{DE}(x_i)$ and $M(x_i)$ can only show up at the
next--to--leading $O(p^4)$ in the chiral expansion.
Of course, $E_B(x_i)$ also  receives corrections from
$\left. A(K \ra \pi_1 \pi_2) \right|_{O(p^4)}$ at this chiral order.

\subsection{\bf $K_{L,S} \ra \pi^0 \pi^0 \gamma$}
For the decays $K^0 \ra \pi^0 \pi^0 \gamma$, Bose statistics implies
\beqa
E(x_2,x_1) &=& - E(x_1,x_2) \no \\
M(x_2,x_1) &=& - M(x_1,x_2)~. \label{Bose}
\eeqa
In the limit where CP is conserved, the amplitude for $K_L$($K_S$) is
purely electric (magnetic).

The transition $K_L \ra \pi^0 \pi^0 \gamma$ has recently been considered
in the literature \cite{FK93,HS93a}.
Eq.~(\ref{Bose}) implies the absence
of a local amplitude of $O(p^4)$, or more generally the absence of an E1
amplitude. Although this by itself does not imply a vanishing one--loop
amplitude (as can be seen in the case of $K_L \ra \pi^+ \pi^- \gamma$
later in this section), Funck and Kambor \cite{FK93} have shown that it
does indeed vanish.

Thus, the decay $K_L \ra \pi^0 \pi^0 \gamma$ is at
least $O(p^6)$ in CHPT. In fact, chiral symmetry permits local octet
couplings of $O(p^6)$ contributing to this transition. A typical
such term, considered in Ref.~\cite{ENPb} with a coupling strength
suggested by chiral dimensional analysis \cite{GM84},
gives rise to a branching ratio
\beq
\left. BR(K_L \ra \pi^0 \pi^0 \gamma) \right|_{O(p^6)} = 7 \cdot 10^{-11}~.
\eeq
By relating $K_L \ra \pi^0 \pi^0 \gamma$ to the decay
$K_L \ra \pi^+ \pi^- \gamma$ (which is dominantly M1), Heiliger and Sehgal
obtain a considerably bigger estimate \cite{HS93a}
$\left. BR(K_L \ra \pi^0 \pi^0 \gamma)\right|_{\rm HS} = 1 \cdot 10^{-8}$
together with
$\left. BR(K_S \ra \pi^0 \pi^0 \gamma)\right|_{\rm HS} = 1.7 \cdot
10^{-11}$.

Recentely an experimental upper bound has been obtained for the
$K_L$-decay: $BR(K_L \ra \pi^0 \pi^0 \gamma)< 10^{-4}$
\cite{E769KP0P0g}.

\subsection{\bf $K_S \ra \pi^+ \pi^- \gamma$}
\label{subsec:KSppg}
\begin{figure}
    \begin{center}
       \setlength{\unitlength}{1truecm}
       \begin{picture}(10.0,3.0)
       \epsfxsize 10. true  cm
       \epsfysize 3.  true cm
       \epsffile{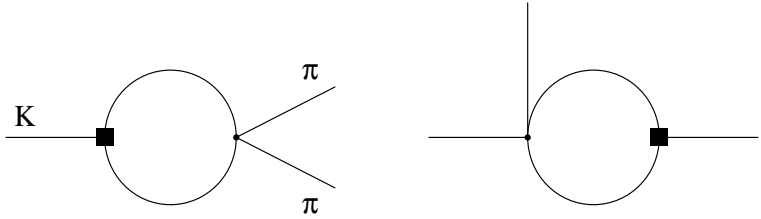}
       \end{picture}
    \end{center}
    \caption{One--loop diagrams for $K \to \pi\pi\gamma$.
The photon has to be attached to any charged line or to any vertex. }
    \protect\label{fig:kppgloops}
\end{figure}
In the limit of CP conservation, the amplitudes for $K_S \ra \pi^+ \pi^-
\gamma$ obey the symmetry relations
\beq
E(x_-,x_+) = E(x_+,x_-)~, \qquad\qquad
M(x_-,x_+) = -M(x_+,x_-)~. \label{KSCP}
\eeq
To $O(p^4)$, the amplitude is therefore purely electric.

According to Sect.~\ref{subsec:Low}, the  amplitude
of $O(p^2)$ is completely determined by bremsstrahlung:
\beq
E_B(x_i) = - \dfrac{e A(K_1^0 \ra \pi^+\pi^-)}{M_K(\dfrac{1}{2}-x_+)
(\dfrac{1}{2}-x_-)}~, \qquad \qquad p_1 = p_+, \quad p_2 = p_-~.
\eeq
At $O(p^4)$, the local contribution to the direct emission amplitude
$E_{DE}(x_i)$ is \cite{DAMS}
\beq
E_4^{\rm local} = -\dfrac{4ie G_8 M_K^3}{F} (N_{14} - N_{15} - N_{16}
- N_{17})~.
\eeq
The combination of coupling constants $N_{14} - N_{15} - N_{16} - N_{17}$
is scale independent \cite{KMWa,EKW} and it also appears in the electric
amplitude for the decay $K^+ \ra \pi^+ \pi^0 \gamma$ \cite{ENPa,ENPb}
(cf. Table \ref{tab:Ni}).
Consequently, the loop amplitudes for both $K_S \ra \pi^+ \pi^- \gamma$
and $K^+ \ra \pi^+ \pi^0 \gamma$ are finite. The (dominant)
pion loop amplitude for $K_S \ra \pi^+ \pi^- \gamma$ (see Ref.~\cite{DAI}
for the complete loop amplitude) is of the form \cite{DAMS,DAI}:
\beq
E_4^{\rm pion~loop} = -\dfrac{ie G_8 M_K^3}{ \pi^2 F} (1-r_{\pi}^2) H(z)~,
\qquad \qquad z=1 - 2x_3~,
\eeq
where the function $H(z)$ is defined in Appendix~\ref{app:loop}.

At present, experimental data \cite{PDG} are consistent with a pure
bremsstrahlung amplitude. However, DA$\Phi$NE
should be able to detect interference with the $O(p^4)$
amplitude that is expected to show up at the level of
$10^{-5}$--$10^{-6}$ in
branching ratio (for $E_\gamma > 20$~MeV) \cite{DAMS,DAI}.

\subsection{\bf $K_L \ra \pi^+ \pi^- \gamma$}
The bremsstrahlung amplitude  \cite{ENPa,ENPb} violates CP:
\beq
E_B(x_i) = - \dfrac{\ve e A(K_1^0 \ra \pi^+\pi^-)}{M_K(\dfrac{1}{2}-x_+)
(\dfrac{1}{2}-x_-)}~, \qquad \qquad p_1 = p_+, \quad p_2 = p_-~. \label{EBREMS}
\eeq
Here $\ve$ is the standard CP violation parameter in
$K \ra \pi\pi$ decays and we have neglected $\ve'$. This suppression of
$E_B(x_i)$ facilitates the experimental observation of the direct emission
amplitude. From $O(p^4)$ on we assume CP conservation implying
\beqa
E(x_-,x_+) &=& -E(x_+,x_-) \no \\
M(x_-,x_+) &=& M(x_+,x_-)~. \label{KLCP}
\eeqa
The dominant contribution of $O(p^4)$ occurs in the magnetic amplitude
and it is due to the anomaly. As discussed in Sect. \ref{subsec:anomaly},
there is no reducible anomalous amplitude of $O(p^4)$. The direct weak
anomaly functional gives rise to \cite{BEP,ENPb,Cheng}
\beq
M_4 = - \dfrac{e G_8 M_K^3}{2\pi^2 F} (a_2 + 2a_4) \label{KLM4}
\eeq
in terms of the coupling constants $a_i$ defined in (\ref{Nan}).

Because of (\ref{KLCP}) there is no local contribution to the electric
amplitude $E(x_i)$ at $O(p^4)$. In contrast to $K_L \ra \pi^0 \pi^0 \gamma$,
there is however a finite one--loop amplitude which turns out
\cite{ENPa,ENPb} to be very small,
\beqa
\left| \dfrac{E_4^{\rm loop}}{E_B} \right| &\leq&
1.2 \cdot 10^{-2} ~.
\eeqa
This small ratio is mainly due to the
antisymmetry in $x_+$, $x_-$ of $E_4^{\rm loop}$ dictated by CP
invariance [cf. Eq.~(\ref{KLCP})].  $O(p^6)$ counterterms could
substantially enhance the direct emission electric amplitude,
which, however, can hardly be detected \cite{DAI}.

There are on the other hand strong experimental
indications for the presence of a sizeable magnetic amplitude beyond
$O(p^4)$. A recent analysis of $K_L \ra \pi^+ \pi^- \gamma$ at FNAL
\cite{E731} confirms an earlier result from Brookhaven \cite{BNL80}
finding evidence for a dependence of the direct emission amplitude on
the photon energy. On the other hand, the dominant direct emission
amplitude $M_4$ in (\ref{KLM4}) is a constant, independent of the photon
energy.

At $O(p^6)$, CP invariance leads to the following most general form
of the magnetic amplitude via Eq.~(\ref{KLCP}):
\beq
M_6(x_+,x_-) = a + b(x_+ + x_-) = a + b - b x_3 = \wh M_6 - b x_3~,
\qquad x_3 = \dfrac{Pq}{M^2_K} = \dfrac{E_\gamma}{M_K}~.
\label{x3}
\eeq
To $O(p^6)$, the total magnetic amplitude is therefore given by
\beq
M(x_+,x_-) = M_4 + M_6(x_+,x_-) = M_4 + \wh M_6 - b x_3 =
(M_4 + \wh M_6)(1 + c x_3)~. \label{defc}
\eeq
 From the distribution in $E_\gamma$ measured by E731 \cite{E731}, one
can extract \cite{Ram} a value
\beq
c = - 1.7 \pm 0.5 \label{cRam}
\eeq
for the slope $c$, in agreement with the earlier measurement
\cite{BNL80,LittV}.

The dominant contributions of $O(p^6)$ in CHPT were analysed in \cite{ENPb}.
First of all, there is a reducible amplitude due to the anomaly of the form
\beq\label{fterm}
M_6^{\rm anom} =  \dfrac{e G_8 M_K^3}{2\pi^2 F} F_1~, \label{M6an}
\eeq
$$
F_1 = \dfrac{1}{1 - r_\pi^2} -
\dfrac{(c_\Theta - \sqrt{2}\,s_\Theta)(c_\Theta + 2\sqrt{2}\,\rho s_\Theta)}
{3(r_\eta^2 - 1)} + \dfrac{(\sqrt{2}\,c_\Theta + s_\Theta)(2\sqrt{2}\,
\rho c_\Theta -s_\Theta)}{3(r_{\eta'}^2 -1)}
$$
$$
r_i = M_i/M_K, \qquad c_\Theta = \cos \Theta, \qquad s_\Theta = \sin \Theta.
$$
In this formula, $\Theta$ denotes the $\eta$--$\eta'$
mixing angle and $\rho \neq 1$ takes into account possible deviations
from nonet symmetry for the non--leptonic weak vertices (nonet symmetry
is assumed for the strong WZW vertices).
At $O(p^4)$ ($\Theta = 0$, $M_{\eta'} \ra \infty$), $F_1$ vanishes
because of the Gell-Mann--Okubo mass formula. In the real world, the
$\eta$ and $\eta'$ contributions interfere destructively for
$0 \leq \rho < 1$ and $\Theta \simeq - 20^\circ$ as in the similar case of
the $K_L \ra 2\gamma$ amplitude (cf. Sect.~\ref{sec:gg}).
Although not really predictable with any
precision, $F_1$ is dominated by the pion pole and certainly positive.

Vector meson exchange also enters at $O(p^6)$. There is a unique
(for $M_\pi = 0$) reducible amplitude originating from the strong VMD
amplitude of $O(p^6)$ via a weak rotation \cite{ENPa,ENPb}:
\beq
M_6^{\rm VMD} = 2 C_V (1 - 3x_3)~,  \qquad
C_V = -\dfrac{16 \sqrt{2} \, e G_8 g_V h_V M_K^5}{3 M_V^2 F} \qquad
(M_V = M_\rho)~.
\eeq
The vector meson couplings $g_V$, $h_V$ \cite{EPRd,EGLPR} are approximately
\beq
g_V \simeq \dfrac{F_\pi}{\sqrt{2}\,M_V}, \qquad
|h_V| \simeq 3.7 \cdot 10^{-2}~.
\eeq
However, there is also a model--dependent direct weak amplitude of $O(p^6)$
related to $V$ exchange. In the factorization model \cite{EKW}
\beq
M_6^{\rm FM} = 4 k_f C_V x_3 - \dfrac{e G_8 M_K^5 L_9}{\pi^2 F^3} k_f
(2 - 5x_3)~,
\eeq
where $k_f$ is a fudge factor which naive factorization puts
equal to one [$k_f = 1$ corresponds to $a_i =1$
$(i = 1,\ldots,4)$ in Eq.~(\ref{Nan})].
Altogether, one obtains \cite{ENPb} for the magnetic amplitude
\beq
M(x_3) = -\dfrac{eG_8 M_K^3}{2\pi^2 F} \left\{ a_2 + 2a_4 - F_1 +
r_V [1 + x_3(2k_f - 3)] + \dfrac{2L_9 M_K^2}{F^2} k_f(2 - 5x_3)\right\}
\label{MKL}
\eeq
$$
r_V = \dfrac{64 \sqrt{2} \,\pi^2 g_V h_V M_K^2}{3 M_V^2} \simeq 0.4
\simeq \dfrac{2 L_9^r(M_\rho) M_K^2}{F_\pi^2}~.
$$

The recent measurement \cite{E731} of the direct emission branching ratio
\beq
BR(E_\gamma > 20\mbox{ MeV})_{\rm DE} = (3.19 \pm 0.16)\cdot 10^{-5}
\label{BRKL}
\eeq
can be used to determine the quantity $a_2 + 2a_4 - F_1$ for given values of
$k_f$. Then, the slope parameter $c$ defined in (\ref{defc}) can
be extracted from Eq.~(\ref{MKL}) both in magnitude and sign.
The values for c obtained by such an analysis \cite{ENPb} are in
agreement with the experimental result (\ref{cRam}).

A compilation and a critical discussion of previous literature on
$K_L \ra \pi^+ \pi^- \gamma$ can be found in Ref.~\cite{ENPb}.

\subsection{$K^+ \ra \pi^+ \pi^0 \gamma$}
The decay $K^+ \ra \pi^+ \pi^0 \gamma$ shares several features with
$K_L \ra \pi^+ \pi^-\gamma$:
\begin{itemize}
\item The bremsstrahlung amplitude is suppressed;
\item The dominating contribution of $O(p^4)$ is due to the chiral anomaly;
\item The one--loop amplitude is finite, but again very small.
\end{itemize}

The bremsstrahlung amplitude \cite{ENPa,ENPb}
is the complete amplitude of $O(p^2)$ and it is suppressed by the
$\Delta I = 1/2$ rule:
\beq
E_B(x_i) = -\dfrac{e A(K^+ \ra \pi^+ \pi^0)}{M_K x_3(\dfrac{1}{2} -x_0)}~,
\qquad \qquad p_1 = p_+~, \quad p_2 = p_0~. \label{EBKP}
\eeq

The magnetic amplitude of $O(p^4)$ consists of both a reducible and a
direct amplitude \cite{ENPa,BEP}:
\beq
M_4 = -\dfrac{e G_8 M_K^3}{2 \pi^2 F} \left(-1 + \dfrac{3}{2} a_2 - 3a_3
\right).\label{M4Kp}
\eeq
Factorization suggests constructive interference between these two
terms.

In contrast to $K_L \ra \pi^+ \pi^- \gamma$, there is now a local
scale--independent contribution of $O(p^4)$ to the electric amplitude
\cite{ENPa,ENPb}:
\beq
E_4^{\rm local} = -\dfrac{2ie G_8 M_K^3}{F} (N_{14} - N_{15} - N_{16}
- N_{17})~. \label{E4KP}
\eeq
As can be seen in Table \ref{tab:Ni}, the same combination of coupling
constants appears in the amplitude for $K_S \ra \pi^+ \pi^- \gamma$
(Sect.~\ref{subsec:KSppg}). By measuring the energy spectrum of the photon,
the counterterm amplitude (\ref{E4KP}) can in principle be isolated through
its interference with the bremsstrahlung amplitude (\ref{EBKP}). One can
estimate the size of this interference by appealing to the factorization
model which predicts \cite{EKW}
\beq
N_{14} - N_{15} - N_{16} - N_{17} = - k_f \dfrac{F_\pi^2}{2 M_V^2} =
- 7 \cdot 10^{-3} k_f~.
\eeq
For $k_f > 0$, the interference is positive
\cite{ENPa,ENPb}:
\beq
\dfrac{E_4^{\rm local}}{E_B} \simeq 2.3 x_3(1 - 2x_0)
(- N_{14} + N_{15} + N_{16} + N_{17})/7 \cdot 10^{-3}~.
\eeq
The sign is well--determined because the ratio $G_8/G_{27}$ is known to be
positive from $K \ra 2\pi$ decays [see Eq.~(\ref{eq:G8G27})]. Except for small
$E_\gamma$ ($x_3 \ra 0$, $2x_0 \ra 1$) where bremsstrahlung is bound to
dominate, the amplitude $E_4^{\rm local}$ should be detectable. In fact,
the experiment of Abrams et al. \cite{KPexp} is consistent with
constructive interference between $E_B$ and $E_4^{\rm local}$,
but the available data are not precise enough
to separate the amplitudes $E - E_B$ and $M$ experimentally.

Since the counterterm amplitude (\ref{E4KP}) is scale
independent, the loop amplitude must be finite.
As in the case of $K_L \ra \pi^+ \pi^- \gamma$, its contribution is
again very small \cite{ENPb,DAI},
\beq
\left| \dfrac{E_4^{\rm loop}}{E_B} \right| \leq 4.4 \cdot 10^{-2} ~.
\eeq
At least in the foreseeable future, the loop amplitude can safely be
neglected in comparison with the bremsstrahlung amplitude (\ref{EBKP}).
On the other hand, the counterterm amplitude (\ref{E4KP})  should be
within reach of experiments at DA$\Phi$NE.

For $K_L \ra \pi^+ \pi^- \gamma$, it was essential to include $V$
exchange effects of $O(p^6)$, in particular to understand the slope
parameter $c$. All the mechanisms discussed there also contribute to
$K^+ \ra \pi^+ \pi^0 \gamma$. Including the $O(p^4)$ amplitude
(\ref{M4Kp}), the total magnetic amplitude is \cite{ENPb}
\beq
M = M_4 + M_6 = -\dfrac{e G_8 M_K^3}{4 \pi^2 F} \left\{ -2 +3a_2 - 6a_3
+ r_V(2k_f -1) + \dfrac{2L_9 M_K^2}{F^2} k_f (3-8x_+ -2x_0)\right\}.
\label{MKP}
\eeq
Under the assumption that direct emission is entirely due to the magnetic
part, experiments \cite{PDG,KPexp} find a branching ratio
\beq
BR(55 < T_{\pi^+}\mbox{(MeV)} < 90) = (1.8 \pm 0.4) \cdot 10^{-5}
\eeq
for the given cuts in the kinetic energy of the charged pion.
Proceeding in a similar way as for $K_L \ra \pi^+ \pi^- \gamma$, the quantity
$A_4 = -2 + 3a_2 - 6a_3$ can be extracted from the measured rate.
For the values of $r_V$ and $L_9$ listed in Eq.~(\ref{MKL}),
$A_4$ is found to be \cite{ENPb}
\beq
A_4 = - 4.1 - 0.3 k_f \pm 0.5~, \qquad 0 \leq k_f \leq 1~.
\eeq

\noindent The following conclusions can be drawn:
\begin{itemize}
\item Compared with $K_L \ra \pi^+ \pi^- \gamma$, the $V$ exchange
contributions are of less importance in the present case. Especially
for $k_f \simeq 1$, the $O(p^6)$ terms are essentially negligible in the
rate.
\item The last term in Eq.~(\ref{MKP}) shows a rather
pronounced dependence on $x_+$. A high--precision analysis of the decay
distribution in $T_{\pi^+}$ may be able to reveal this dependence.
\item The fitted values of $A_4$ are very much consistent with our
expectations based on $a_i \; \leq \; 1$ (cf. Sect. \ref{subsec:anomaly}).
\item Because of the expected positive interference between $E_B$
and $E_4^{\rm local}$, the coefficient $|A_4|$ is probably somewhat
smaller than found above. Future experimental analysis at DA$\Phi$NE should
include an E1 amplitude of the type (\ref{E4KP}).
\end{itemize}

\subsection{CP violation}

The charge asymmetry in $K^\pm \to \pi^\pm\pi^0 \gamma$
is a very promising observable for the detection of direct CP violation
in radiative kaon decays. Although suppressed by
the electromagnetic coupling, this  channel
may be competitive with $K\to \pi\pi$ since it is not
suppressed by the $\Delta I=1/2$ rule.
CP violation in $K^\pm \to \pi^\pm \pi^0 \gamma$ arises, as in the
$K^\pm \to \pi^\pm \gamma \gamma$ case, from the interference between
the absorptive part of the electric amplitude (starting only at
$O(p^4)$) and the imaginary part of the counterterm contribution
(\ref{E4KP}). As in the case of
$K^\pm \to \pi^\pm \gamma \gamma$, the imaginary part of the weak
counterterm is generated at short distances.

The authors of Ref.~\cite{DibP} claimed that the electric dipole operator
could give a large contribution to the imaginary part of the weak
counterterms, obtaining the following estimate:
\beq
{\Delta \Gamma (K^\pm \rightarrow \pi^\pm \pi^0 \gamma) \over
2 \Gamma(K^\pm \rightarrow \pi^\pm \pi^0\gamma)} \le 1 \cdot 10^{-3}~.
\label{eq:cpvioldp}
\eeq
This estimate seems to be too optimistic since the counterterm introduced in
Ref.~\cite{DibP} is of O$(p^6)$ and does not have the correct chiral
suppression factor. A more realistic estimate for the charge asymmetry
is $\Delta \Gamma /2 \Gamma \sim 10^{-5}$  \cite{PaverR}. In any case,
also with the large estimate of Eq.~(\ref{eq:cpvioldp}),
DA$\Phi$NE will not have  enough statistics to establish
$\Delta \Gamma \neq 0$ and will only put an
interesting upper limit on the asymmetry.

A possible goal for DA$\Phi$NE could be to improve the measurement of
\beq
\eta_{+-\gamma}={{A(K_L \ra \pi^+ \pi^- \gamma)_{IB+E1}}\over{A(K_S \ra \pi^+
\pi^- \gamma)_{IB+E1}}}, \label{eta+-gamma}
\eeq
which has already been measured at Fermilab \cite{etapmga, etapmgb}.
The result for the IB contribution is \cite{etapmgb}:
\beq
\vert\eta_{+-\gamma (IB)}\vert=
\Big\vert \frac{A( K_L\rightarrow \pi^+\pi^-\gamma)_{IB}}
{A(K_S\rightarrow \pi^+ \pi^-\gamma)_{IB}}\Big\vert
 = (2.414 \pm 0.065 \pm 0.062 )\cdot 10^{-3}~,\label{etagamma}
\eeq
\beq
\phi_{+-\gamma (IB)}=\arg (\eta_{+-\gamma (IB)})
 = (45.47 \pm 3.61 \pm 2.40)^\circ~.
\label{phigamma}
\eeq
DA$\Phi$NE \ should be able to improve these values studying the time
evolution of the decay \cite{NG,Pugliese}.

\subsection{Improvements at DA$\Phi$NE}

The detection of both charged and neutral $K\to\pi\pi\gamma$
decays with high statistics in the same experimental setup will lead to
several interesting tests of CHPT.

The measurements of the electric direct emission
components of $K_S \to \pi^+ \pi^- \gamma$ and
$K^+ \to \pi^+\pi^0 \gamma$, where the same counterterm combination
appears, will furnish a consistency check of the $O(p^4)$
contributions. Moreover, the measurements of the photon energy spectrum in
$K_L \ra \pi^+ \pi^- \gamma$ will lead to a substantial improvement in the
determination of the slope parameter which is sensitive to
contributions of $O(p^6$). This will allow for checks of theoretical
models.

\setcounter{equation}{0}
\setcounter{subsection}{0}

\section{Kaon decays with a lepton pair in the final state}
\label{sec:Dalitz}

We confine the discussion to final states with charged leptons only.
Decays with a neutrino pair in the final state are of great theoretical
interest and we refer to Ref.~\cite{reviews} for detailed reviews
of both theoretical and experimental aspects. However,
DA$\Phi$NE will not be able to gather significant statistics for
$K^+\to \pi^+ \nu \bar{\nu}$ with a branching ratio around $10^{-10}$,
not to speak of the CP violating process $K_L\to \pi^0 \nu \bar{\nu}$.

The decays with a Dalitz pair are in general dominated by one--photon
exchange. This is certainly the case for transitions that occur already
for a real photon like $K^0\to l^+ l^- \gamma$, but also for
many decays that vanish for $q^2=0$ like $K\to \pi l^+ l^-$ or
$K_L\to \pi^0 \pi^0 l^+ l^-$. The decays $K^0\to l^+ l^-$, on the other
hand, are dominated by two--photon exchange.
For the consideration of CP violating effects, genuine short--distance
mechanisms like $Z$--penguin and $W$--box diagrams must be taken
into account. However, experiments at DA$\Phi$NE
will only be able to give limits for CP violating observables
for the class of decays considered in this section.

\subsection{No pions in the final state}
\label{sec:nopion}
\subsubsection{$K^0\to \gamma^* \gamma^*$}

For a transition
$$
M(p) \ra \gamma^*(q_1) + \gamma^*(q_2), \qquad p^2 = M^2~,
$$
gauge invariance implies the following general decomposition of the
amplitude \cite{EPRc}:
\beqa
M^{\mu\nu}(q_1,q_2) &=&
\left(g^{\mu\nu} -
\dfrac{q_1^\mu q_1^\nu}{q_1^2} - \dfrac{q_2^\mu q_2^\nu}{q_2^2} +
\dfrac{q_1 \cdot q_2}{q_1^2 q_2^2} \; q_1^\mu q_2^\nu \right)
M^2 a(q_1^2,q_2^2) \no \\
&& \mbox{} + \left[q_2^\mu q_1^\nu - q_1 \cdot q_2 \left(
\dfrac{q_1^\mu q_1^\nu}{q_1^2} + \dfrac{q_2^\mu q_2^\nu}{q_2^2} -
\dfrac{q_1 \cdot q_2}{q_1^2 q_2^2} \; q_1^\mu q_2^\nu \right)\right]
b(q_1^2,q_2^2) \no \\
&& \mbox{} + \ve^{\mu\nu\rho\sigma} q_{1\rho} q_{2\sigma}
c (q_1^2,q_2^2)~.\label{eq:Mg*g*}
\eeqa
Due to Bose symmetry, the invariant amplitudes $a,b,c$ are
symmetric functions of $q_1^2,q_2^2$. With CP conserved, $a$ and $b$
contribute to $K_1^0 \ra \gamma^* \gamma^*$ while $c$ contributes
to $K_2^0 \ra \gamma^* \gamma^*$.

When one of the photons is on--shell $(q_1^2 = 0)$, $M^{\mu\nu}$ is
described by two invariant amplitudes:
\beq
M^{\mu\nu}(q_1,q_2) = (q_2^\mu q_1^\nu - q_1 \cdot q_2 g^{\mu\nu})
b(0,q_2^2) + \ve^{\mu\nu\rho\sigma} q_{1\rho} q_{2\sigma}
c(0,q_2^2)~.
\eeq
The decay widths are
\beqa
\Gamma(K_S \ra \gamma \gamma) &=& \dfrac{M_K^3}{32 \pi} |b(0,0)|^2 \\
\Gamma(K_L \ra \gamma \gamma) &=& \dfrac{M_K^3}{64 \pi} |c(0,0)|^2 \no
\eeqa
when both photons are on--shell.

In terms of these widths for the on--shell decays, the differential
rates for  $K_{L,S}\to \gamma l^+ l^-$ can be written
\beq
{{\rm d}\Gamma(K_{L(S)}\to \gamma l^+ l^-) \over {\rm d}z}
= \Gamma(K_{L(S)}\to \gamma \gamma) |R_{L(S)}(z)|^2 {2\alpha \over 3\pi}
{(1-z)^3 \over z} \left(1+{2r_l^2 \over z}\right)
\sqrt{1-{4r_l^2 \over z}}~,\label{eq:llg}
\eeq
where
$$ z={q_2^2 \over M_K^2}~, \qquad   r_l={m_l \over M_K} \qquad
( 4r_l^2 \le z \le 1)$$
and $R(z)$ is a form factor normalized to $R(0)=1$.

\vspace*{.6cm}
\centerline{$K_S\to \gamma l^+l^-$}
\vspace*{.4cm}

\noindent
The form factor $R_S(z)$ was calculated to $O(p^4)$ in CHPT \cite{EPRc}:
\beq
R_S(z) = H(z)/H(0)
\eeq
where the loop function $H(z)$ can be found in Appendix~\ref{app:loop}.
The corresponding decay rates for the Dalitz pair modes normalized to
$\Gamma(K_S \ra \gamma \gamma)$ are compared in Table
\ref{tab:dalitz} with phase space and with a dispersion model
\cite{SE73}. Only for the muon channel could the predictions of the
Standard Model to $O(p^4)$  in principle be discriminated. The same
qualification applies to the spectra given by Eq.~(\ref{eq:llg}).
Therefore, DA$\Phi$NE will not distinguish between phase space
and the CHPT prediction, but it should permit detection of the electronic
mode.
\begin{table}
\label{tab:dalitz}
$$
\begin{tabular}{|l|c|c|} \hline
& $l = e$ & $l = \mu$ \\ \hline
Standard Model to $O(p^4)$ & $1.60 \cdot 10^{-2}$ & $3.75 \cdot 10^{-4}$ \\
dispersion model \cite{SE73} & $1.59 \cdot 10^{-2}$ &
$3.30 \cdot 10^{-4}$ \\
phase space & $1.59 \cdot 10^{-2}$ & $4.09 \cdot 10^{-4}$ \\ \hline
\end{tabular}
$$
\caption{Normalized rates $\Gamma(K_S \ra \gamma l^+ l^-)/\Gamma
(K_S \ra \gamma \gamma)$}
\end{table}

\vspace*{.6cm}
\centerline{$K_L\to \gamma l^+l^-$}
\vspace*{.4cm}

\noindent
For the $K_L$ decays we face the problem already
encountered in Sect.~\ref{sec:gg}, namely, that it is difficult to
obtain a reliable prediction for $K_L\to \gamma\gamma$ in CHPT.
The amplitude is formally of $O(p^6)$ in the
chiral expansion, although it is numerically more like a typical $O(p\,^4)$
amplitude. Because the cancellation at $O(p^4)$ involves two rather big
amplitudes, $A(K_L \ra \gamma\gamma)$ is very sensitive to the chiral
corrections which include in particular
$\eta - \eta\,'$ mixing. Since there is at this time no complete calculation
of $K_L \ra \gamma\gamma$ in CHPT to $O(p^6)$, we will instead
use the experimental value of $|A(K_L \ra \gamma\gamma)|$ whenever
needed in this section.

The reducible amplitude is given by
a VMD contribution to the transitions $P^0 \ra \gamma l^+ l^-$
$(P^0 = \pi^0,\eta_8,\eta_1)$ supplemented by an
external weak transition $P^0 \ra K_L$ induced by ${\cal L}_2^{\Delta S=1}$.
To first order in $z$, one obtains \cite{Gosen}
\beq
R_L^{VMD}(z) = 1 + c_V {M_K^2\over M_V^2} z~,  \qquad M_V=M_\rho
\eeq
where the coefficient
\beq
c_V = {32 \sqrt{2} \pi^2 f_V h_V\over 3}
\eeq
contains the chiral vector meson couplings $f_V$, $h_V$
\cite{EGLPR,EPRd}. Its absolute value is $|c_V| = 0.94$ in agreement with
the analysis of Ref.~\cite{BBC90} for $P^0 \ra \gamma\gamma^*$ and with
experiment, which requires in addition $c_V > 0$. However, there is as usual
also a direct weak contribution of $O(p^6)$ to $K_L \ra \gamma\gamma^*$
which is model dependent. The total form factor to first
order in z is \cite{Gosen}
\beq
R_L(z) = 1 + (c_V + c_D)
{M_K^2\over M_V^2} z
\eeq
with a direct weak coefficient $c_D$. Choosing the relative sign
between $c_D$ and $c_V$ appropriately, the weak deformation
model \cite{EPRd} predicts
$$
b=0.7
$$
for the slope $b$ defined via
\beq
R_L(z) = 1 + b z + O(z^2)~,\label{eq:slope}
\eeq
in agreement with the experimental result \cite{Barr90a}
\beq
b_{exp} = 0.67 \pm 0.11~.\label{eq:bexp}
\eeq
Thus, there is definitely experimental evidence for a direct weak amplitude
interfering constructively with the VMD amplitude. A direct amplitude was
already put forward in Ref.~\cite{BMS83} in terms of a
$\gamma^* - K^*$ transition. Their form factor has the form
\beq
R_L(z)= {1 \over 1-0.418z}+{c \alpha_{K^*}\over 1-0.311z}
\left[{4\over 3}-{1\over 1-0.418z}-{1\over 9(1-0.405z)}-{2\over 9(1-0.238z)}
\right] \label{eq:BMS}
\eeq
where $c$ was evaluated in terms of known coupling constants to be
$c=2.5$ and $\alpha_{K^*}$ parametrizes the unknown electroweak transition
$\gamma^* - K^*$.
In the vacuum insertion approximation the authors of Ref.~\cite{BMS83}
obtained $|\alpha_{K^*}|\simeq 0.2 \sim 0.3$. The experimental value
(\ref{eq:bexp}) corresponds to $\alpha_{K^*}= - 0.28\pm 0.12$.

Once the slope is given, one can use (\ref{eq:slope}) in (\ref{eq:llg})
to predict the rate. The agreement with the experimental results
\beq
BR(K_L\to e^+ e^- \gamma)_{\rm exp} = \left\{
\ba{lr} (9.2\pm 0.5 \pm 0.5)\cdot 10^{-6}& \cite{Barr90a}\\
(9.1\pm 0.4~^{+0.6}_{-0.5})\cdot 10^{-6} & \cite{Ohl90}\ea \right.
\eeq
confirms that the linear approximation for $R_L(z)$ is sufficient.
DA$\Phi$NE will have more statistics for  $K_L\to \gamma e^+e^-$ and
will also be able  to measure  $K_L\to \gamma \mu^+\mu^- $, where the
dependence on the form factor is stronger. Both with a linear
form factor (\ref{eq:slope}) and with the VMD form factor (\ref{eq:BMS})
one arrives at essentially the same prediction \cite{Gosen,BMS83}
\beq
{\Gamma( K_L\to \gamma e^+e^-) \over \Gamma( K_L\to \gamma \mu^+\mu^-)}
\simeq 25~.
\eeq

\vspace*{.6cm}
\centerline{$K_L\to e^+e^-e^+e^-$}
\vspace*{.4cm}

\noindent
The decay $K_L\to e^+e^-e^+e^-$ is sensitive to the invariant amplitude
$c(q_1^2,q_2^2)$ in Eq.~(\ref{eq:Mg*g*}). There is no
explicit theoretical calculation yet.
DA$\Phi$NE will improve the existing value for the branching ratio
\beq
BR(K_L\to e^+ e^- e^+ e^-)_{\rm exp} =
 (3.9\pm 0.7 )\cdot 10^{-8} ~\cite{PDG}
\eeq
and will test the $K_L\to \gamma^*\gamma^*$ amplitude.

On the basis of Table \ref{tab:dalitz}, DA$\Phi$NE will not be able
to see the corresponding decay $K_S\to e^+ e^- e^+ e^-$.

\subsubsection{\bf $K^0\to l^+ l^-$}
\label{sub:ll}
The amplitude for the decay of a $K^0$ into a lepton
pair has the general form
\beq
A(K^0 \ra l^+ l^-) = \bar u(iB + A\gamma_5) v
\eeq
with a decay rate
\beq
\Gamma(K^0 \ra l^+ l^-) = \dfrac{M_K\beta_l}{8\pi}
(|A|^2 + \beta^2_l |B|^2), \qquad
\beta_l = \left( 1 - \dfrac{4 m_l^2}{M_K^2} \right)^{1/2}.
\eeq
In the CP limit, the transition
$K_1^0 \ra l^+ l^-$ is determined by the $p$--wave amplitude $B_1$
while $K_2^0 \ra l^+ l^-$ proceeds only via the $s$--wave $A_2$.

There is a unique lowest--order coupling \cite{EP91}
\beq
\partial^\mu K_2^0 \bar \psi \gamma_\mu \gamma_5 \psi \label{eq:Kll}
\eeq
that contributes only to the $s$--wave amplitude $A_2$. There is in fact
a well--known short--distance amplitude of this type for
$K_L \ra l^+ l^-$ \cite{reviews} with an important top--quark contribution.
However, this decay is dominated by the absorptive part of the transition
$K_L \ra \gamma^* \gamma^* \ra l^+ l^-$~:
\beq
|\mbox{Im } A_2^{\gamma\gamma}| = \dfrac{\alpha m_\mu}{4\beta_\mu M_K}
\ln \dfrac{1 + \beta_\mu}{1 - \beta_\mu}
\left[ \dfrac{64\pi \Gamma(K_L \ra 2\gamma)}{M_K}\right]^{1/2}
\eeq
or
\beq
BR(K_L \ra \mu^+ \mu^-)_{\rm abs}
\simeq \dfrac{\alpha^2 m_\mu^2}{2\beta_\mu M_K^2}
\left(\ln \dfrac{1 + \beta_\mu}{1 - \beta_\mu} \right)^2 B(K_L \ra 2\gamma)~.
\eeq
The experimental branching ratio \cite{PDG} $BR(K_L \ra 2\gamma) =
(5.73 \pm 0.27) \cdot 10^{-4}$ yields
\beq
|\mbox{Im } A_2^{\gamma\gamma}| = (2.21 \pm 0.05) \cdot 10^{-12}
\eeq
and
\beq
BR(K_L \ra \mu^+ \mu^-)_{\rm abs} = (6.85 \pm 0.32) \cdot 10^{-9}~.
\eeq
Comparing with the latest measurements
\beq
BR(K_L \ra \mu^+ \mu^-) = \left\{ \ba{lr}
(7.9\pm 0.6 \pm 0.3) \cdot 10^{-9} & \mbox{ KEK-137} ~\cite{KEK137}\\
(7.0\pm 0.5)\cdot 10^{-9} & \mbox{ BNL-E791} ~\cite{AGS791}\\
(7.4\pm 0.4)\cdot 10^{-9} & \mbox{ PDG} ~\cite{PDG}\ea \right.~,
\eeq
one finds that the absorptive part (unitarity bound) nearly saturates
the total rate.

The dispersive part for the two--photon intermediate state is
model dependent. There are various models in the literature \cite{BMS83,KLll}
that make different predictions for the dispersive part. As
long as not even the sign of the interference between the dispersive
and the short--distance part is established, the decay $K_L\to \mu^+\mu^-$
is of limited use for the determination of the CKM mixing matrix
element $V_{td}$. Because of the
helicity suppression, DA$\Phi$NE will not be able to see the decay
$K_L\to e^+ e^-$ [the unitarity bound due to the absorptive part
is $BR(K_L\to e^+ e^-)= 3\cdot 10^{-12}$],
unless there are contributions from mechanisms beyond
the Standard Model.

The same statement applies to the decays $K_S\to l^+ l^-$. These
decays are theoretically interesting because the lowest--order
amplitude in CHPT is unambiguously calculable \cite{EP91}. It is
given by a two--loop diagram describing the transition
$K_1^0 \ra \gamma^* \gamma^* \ra l^+ l^-$. The
subprocess $K_1^0 \ra \gamma^* \gamma^*$ is determined by the one--loop
diagrams in Fig.~\ref{fig:ksloops} relevant for $K_S\to \gamma\gamma$ as
discussed in Sect.~\ref{sec:gg}.

Normalizing to the rate for $K_S\to \gamma\gamma$, one obtains
the relative branching ratios \cite{EP91}
\beq
\dfrac{\Gamma(K_S \ra \mu^+ \mu^-)}{\Gamma(K_S \ra \gamma \gamma)} =
2 \cdot 10^{-6}~, \qquad \qquad
\dfrac{\Gamma(K_S \ra e^+ e^-)}{\Gamma(K_S \ra \gamma \gamma)} =
8 \cdot 10^{-9} ~,
\eeq
well below the present experimental upper limits \cite{PDG} and
beyond the reach of DA$\Phi$NE.

\subsection{$K\to \pi l^+ l^-$}
\label{subsec:pig}

$K\to \pi \gamma$ with a real photon is forbidden by
gauge and Lorentz invariance. The CP conserving Dalitz pair decays $K^+\to
\pi^+ l^+ l^-$ and $K^0_1\to \pi^0 l^+ l^-$ are dominated by
virtual photon exchange. In accordance with the general theorem
(\ref{eq:pig*}), the leading amplitudes for these transitions
are of $O(p^4)$ in CHPT. At this order, there are both one--loop
contributions and tree--level contributions involving the low--energy
constants $N_{14}$ and $N_{15}$ \cite{EPRa}.
The determination of these coupling constants will be one of the
goals of experiments at DA$\Phi$NE.

The amplitude for
\beq
K(p)\to \pi(p') +\gamma^*(q) \to \pi(p^\prime) l^+(k^\prime) l^-(k) ~,
\qquad    q= k + k^\prime
\eeq
has the general form compatible with Lorentz and gauge invariance
\beq
A(K\to \pi\gamma^* \to \pi l^+ l^-)=-{G_8 e^2 \over (4\pi)^2(q^2+i\epsilon)}
V_\mu(p,q)\bar{u}(k)\gamma^\mu v(k')~,\label{eq:gexch}
\eeq
where
\beq
 V_\mu(p,q)=[q^2(p+p')_\mu-(M_K^2-M_\pi^2)q_\mu] V(z)~,
\qquad z={q^2\over M_K^2}~.\label{eq:Vmu}
\eeq
Within the one--photon exchange approximation, all the dynamics is contained
in the invariant functions $V_+(z)$ and $V_S(z)$ for the decays
$K^+\to\pi^+ l^+ l^-$ and $K_S\to\pi^0 l^+ l^-$, respectively.
The differential decay rate is given by
\beq
{{\rm d}\Gamma\over {\rm d}z}={G_8^2 \alpha^2 M_K^5\over
 192\pi^5}\lambda^{3/2}(1,z,r_\pi^2) (1-{4r_l^2\over z^2})^{1/2}
 (1+{2r_l^2\over z})|V(z)|^2 \label{eq:dGamma}
\eeq
$$
r_\pi={M_\pi\over M_K}~, \qquad r_l={m_l\over M_K}~, \qquad
4 r_l^2 \leq z \leq (1- r_\pi)^2 ~.
$$
The kinematical function $\lambda(x,y,z)$ is defined in
Appendix~\ref{app:loop}.

At $O(p^4)$ in CHPT, the invariant amplitudes are \cite{EPRa}
\beqa
V_+(z) &=& -\vp(z) -\vp(z/r_\pi^2) - w_+ \no\\
V_S(z) &=& 2 \vp(z) + w_S~.\label{eq:V+S}
\eeqa
The loop function $\vp(z)$ can be found in Appendix~\ref{app:loop}. The
scale--independent constants $w_+$ and $w_S$ contain both strong
and weak low--energy constants:
\beqa
w_+ &=& {4\over 3}(4 \pi)^2 \left(N^r_{14}(\mu) - N^r_{15}(\mu)
+ 3 L^r_9(\mu)\right) - {1\over 3} \ln{M_K M_\pi\over \mu^2} \no\\
w_S &=& {2\over 3}(4 \pi)^2 \left(2 N^r_{14}(\mu) + N^r_{15}(\mu)
\right) - {1\over 3} \ln{M_K^2\over \mu^2}~.
\eeqa
Note that the coupling constant $N_{14}$ contains \cite{EPRa,BP93}
the electromagnetic penguin contribution \cite{GW80}.
The branching ratios are quadratic functions of these constants
\cite{EPRa,LittV}:
\beqa
BR(K^+\to \pi^+ e^+ e^-) &=& (3.15 -21.1 w_+ + 36.1 w_+^2)\cdot 10^{-8}
|G_8/9\cdot 10^{-6} {\rm GeV}^{-2}|^2 \no\\
BR(K^+\to \pi^+ \mu^+ \mu^-) &=& (3.93 -32.7 w_+ + 70.5 w_+^2)\cdot 10^{-9}
|G_8/9\cdot 10^{-6} {\rm GeV}^{-2}|^2 \no\\
BR(K_S\to \pi^0 e^+ e^-) &=& (3.07 -18.7 w_S + 28.4 w_S^2)\cdot 10^{-10}
|G_8/9\cdot 10^{-6} {\rm GeV}^{-2}|^2 \no\\
BR(K_S\to \pi^0 \mu^+ \mu^-) &=& (6.29 -38.9 w_S + 60.1 w_S^2)\cdot 10^{-11}
|G_8/9\cdot 10^{-6} {\rm GeV}^{-2}|^2 ~.\label{eq:Kpg}
\eeqa

So far, only the decay $K^+\to \pi^+ e^+ e^-$ has been observed.
A recent experiment at Brookhaven \cite{Alliegro} has reported both a
branching ratio
\beq
BR(K^+\to \pi^+ e^+ e^-)= (2.99\pm 0.22)\cdot 10^{-7}
\eeq
and a value
\beq
w_+= 0.89^{+0.24}_{-0.14} \label{eq:w+}
\eeq
for the constant $w_+$ from a fit to the spectrum shape. In principle,
the branching ratio $BR(K^+\to \pi^+ e^+ e^-)$ depends very sensitively
on $w_+$. However, due to the considerable uncertainty in the octet coupling
$G_8$, the shape of the $z$ distribution is at present a safer
observable to extract $w_+$. Nevertheless, for the central value
of $G_8$ in Eq.~(\ref{eq:G827}) the value for $w_+$ extracted
from the rate agrees with (\ref{eq:w+}) within 1.5$\sigma$ \cite{Alliegro}.

It was pointed out in Ref.~\cite{SW90} that the observation of a
parity violating asymmetry in these decays would reveal the presence
of a short--distance contribution ($Z$--penguin and $W$--box diagrams)
interfering with the dominant one--photon exchange amplitude
(\ref{eq:gexch}). The asymmetry could yield information on the CKM
matrix element $V_{td}$. From the analysis of Ref.~\cite{SW90} one
concludes that DA$\Phi$NE will only be able to place an upper limit
on this asymmetry.

Since $w_+$ and $w_S$ depend on two different combinations of
the weak low--energy constants $N_{14}$ and $N_{15}$, chiral symmetry
does not yield any relation between them. Consequently, chiral
symmetry alone
does not relate the decay amplitudes for $K^+\to \pi^+ l^+  l^-$
and $K_S\to \pi^0 l^+ l^-$. Starting with Ref.~\cite{EPRa}, several
model assumptions have been proposed that would allow us to express
$w_S$ in terms of $w_+$ \cite{EPRd,Cheng,EKW,BP93}. As an
example, we consider the factorization model \cite{EKW} that predicts such
a relation in terms of the parameter $k_f$ encountered previously
($0 < k_f\lsim 1$):
\beq
w_S = w_+ + {2(4\pi F_\pi)^2\over M_V^2}(2 k_f -1) + {1\over 3}
\ln{M_\pi\over M_K} = w_+ + 4.6 (2 k_f -1) - 0.43 ~. \label{eq:wS+}
\eeq
The original model of Ref.~\cite{EPRa} had $k_f=1/2$ as also predicted
by the weak deformation model \cite{EPRd}. As Eq.~(\ref{eq:wS+})
indicates, $w_S$ is extremely sensitive to small deviations from
$k_f = 1/2$ even within the specific factorization model. Moreover,
for the experimental value (\ref{eq:w+}) of $w_+$ the relation
(\ref{eq:wS+}) gives $w_S\simeq 0.5$ for $k_f=1/2$. In the invariant
amplitude $V_S(z)$ in (\ref{eq:V+S}), the function $\vp(z)$
due to the kaon loop varies very little over the physical region
and is approximately given by $\vp(0)= - 1/6$. Thus, there is a strong
destructive interference between the loop contribution and $w_S$
in this case and the resulting branching ratio \cite{EPRa}
$BR(K_S\to \pi^0 e^+ e^-)\simeq 5\cdot 10^{-10}$ is very small. Both in
view of the strong model dependence and of possible higher--order chiral
corrections, such a low rate should be interpreted as an
approximate lower limit rather than as a reliable prediction. To emphasize
this point, we also give the predictions of Ref.~\cite{BP93}
for the two constants $w_+$, $w_S$:
\beqa
w_+ &=& 1.0^{+0.8}_{-0.4} \no\\
w_S &=& 1.0^{+1.0}_{-0.6}~.
\eeqa
It is evident that theory cannot make a reliable prediction
for $w_S$ at this time. On the other hand, DA$\Phi$NE should be able
either to detect the decay $K_S\to \pi^0 e^+ e^-$ or to give at least
a non--trivial upper bound for $|w_S|$.

\subsection{\bf $K_L\to \pi^0 \pi^0 e^+ e^-$}
If there is a non--vanishing $O(p^4)$ amplitude for a process with
a real photon, the corresponding Dalitz pair transition is usually
dominated by the same mechanism. A typical example is the decay
$K_L \ra \pi^+ \pi^- e^+ e^-$ \cite{SW92,HS93b}.
As discussed in Sect.~\ref{sec:1g}, the amplitude for
$K_L\to \pi^+\pi^-\gamma$ is mainly given by
bremsstrahlung and by the magnetic amplitude $M_4$ due to the chiral
anomaly. It has been argued in Refs.~\cite{SW92,HS93b}
that the additional loop contributions, which vanish for a real photon,
are negligible for the Dalitz pair decay amplitude.
The final result for the branching ratio is \cite{HS93b}
\beq
BR(K_L \ra \pi^+ \pi^- e^+ e^-) = 2.8 \cdot 10^{-7}~. \label{eq:HSW}
\eeq

The situation is different for a transition like
$K_L \ra \pi^0 \pi^0 l^+ l^-$ where the amplitude
vanishes to $O(p^4)$ for a real photon, but where a non--zero amplitude
exists for virtual photons: there is a non--vanishing electric amplitude
of $O(p^4)$ to which both counterterms and loop diagrams contribute
\cite{FK93}.

As can be seen from Table \ref{tab:Ni}, the loop amplitude is divergent
and the same combination of counterterms appears as in the decay
$K_S \ra \pi^0 l^+ l^-$ discussed previously.
The amplitude has the form
\beq
A(K_L \ra \pi^0 \pi^0 l^+ l^-) = \Phi(q^2,p \cdot q)
[p \cdot q q^\mu - q^2 p^\mu] \dfrac{e}{q^2} \; \bar u(q_1) \gamma_\mu
v(q_2)
\eeq
where $p = p_1 + p_2$ is the sum of the pion momenta and
$q = q_1 + q_2$ is the virtual photon momentum. The invariant amplitude
$\Phi(q^2,p \cdot q)$ was calculated by Funck and Kambor
\cite{FK93}:
\beqa
\Phi(q^2,p \cdot q) &=& \dfrac{G_8 e}{16 \pi^2 F}
\left\{ \dfrac{2(M_K^2 - M_\pi^2)}{2p \cdot q + q^2} [ \vp (z) -
\vp (\dfrac{z}{r^2_\pi}) + {1\over 3} \ln r_\pi ]  \right. \no \\
&& \left. \mbox{} + 2 \vp (\dfrac{z}{r^2_\pi})  - {2\over 3} \ln r_\pi
+ w_S \right\} \label{eq:FKll}
\eeqa
$$
r_\pi = \dfrac{M_\pi}{M_K} , \qquad z = \dfrac{q^2}{M_K^2}
$$
$$
w_S = \dfrac{32 \pi^2}{3} [ 2 N^r_{14}(\mu) + N^r_{15}(\mu)] -
\dfrac{1}{3} \ln \dfrac{M_K^2}{\mu^2}~.
$$
The function $\vp(z)$ can be found in Appendix~\ref{app:loop}.

The authors of Ref.~\cite{FK93} have made a careful study of the total
rate and of various spectra for different values of $w_S$. The situation
is similar to $K_S\to \pi^0 e^+ e^-$ discussed in Sect.~\ref{subsec:pig}.
The rate is again very sensitive to the coupling constant $w_S$.
The counterterm amplitude interferes constructively
(destructively) with the loop amplitude for negative (positive) $w_S$.
However, even with constructive interference the branching ratio
is at most $10^{-9}$. DA$\Phi$NE will probably only set an upper
limit for the rate.

\subsection{CP violation}
The possibility of studying CP violation
in $K_L \to \pi^+ \pi^- e^+ e^-$  has been
investigated recently \cite{SW92,HS93b}. The employed model contains
(i) a CP conserving amplitude associated with the M1 transition in
$K_L \to \pi^+ \pi^- \gamma$, (ii) an indirect CP violating amplitude related
to the bremsstrahlung part of $K_L \to \pi^+ \pi^- \gamma$, and (iii)
a direct CP violating term  associated with the short--distance interaction
$s\bar{d}\ra e^+ e^-$. Interference between the first two components generates
a large CP violating asymmetry ($\sim 14\%$) in
the angle $\phi$ between the planes of $e^+ e^-$ and $\pi^+\pi^-$.
With the branching ratio
(\ref{eq:HSW}), DA$\Phi$NE will measure the indirect CP violating
contribution in this channel. The effects of direct CP violation
are much smaller and not accessible for DA$\Phi$NE.

Other CP violating observables involving lepton pairs, like the decays
$K_L\to \pi^0 e^+ e^-$ and $K_L\to \pi^0 \nu \bar{\nu}$ \cite{Litt},
the charge asymmetry in $K^\pm\to \pi^\pm e^+ e^-$ and the
transverse muon polarization in $K_L\to \pi^0 \mu^+ \mu^-$ \cite{EPRc}
are beyond the reach of DA$\Phi$NE.

\subsection{Improvements at DA$\Phi$NE}
DA$\Phi$NE should be able to improve the experimental information
about the form factors in the decay $K_L \to e^+ e^-\gamma$. Those
measurements will serve as a test of different model predictions.
One should also obtain a first measurement of  $K_L \to \mu^+ \mu^-\gamma$.
However, due to the smaller number of events, it will be more
difficult to discriminate between different models in this case.

On the other hand, statistics at DA$\Phi$NE should be sufficient
to improve the measurement of the width and of the lepton spectrum
in $K^+ \to \pi^+ e^+ e^-$ and to detect the decay $K^+ \to \pi^+
\mu^+ \mu^-$.

A very interesting channel is $K_S \to \pi^0 e^+ e^-$,
which is extremely sensitive to the counterterm amplitude.
DA$\Phi$NE should be able to detect this decay and to discriminate
among models for the counterterm coupling constants. This
information will be especially important for the indirect CP
violating contribution to $K_L\to \pi^0 e^+ e^-$ \cite{Gabbiani}.

\newpage
\begin{center}
{\large \bf Acknowledgements}
 \end{center}
\noindent
We want to thank our collaborators and the
members of the DA$\Phi$NE working groups for their contributions
and for their helpful comments, especially F. Buccella, L. Cappiello,
A.G. Cohen, D. Espriu, J. Kambor, L. Maiani,
 M. Miragliuolo, N. Paver, A. Pich, A. Pugliese,
E. de Rafael and F. Sannino.
G.D. thanks the CERN Theory Division, where part of this work was done,
for hospitality.

\medskip\medskip
\newcounter{zahler}
\renewcommand{\thesection}{\Alph{zahler}}
\renewcommand{\theequation}{\Alph{zahler}.\arabic{equation}}
\setcounter{zahler}{1}
\setcounter{equation}{0}

\vspace*{1cm}

\section{Loop functions}
\label{app:loop}

The following functions occur in one--loop amplitudes for
the $K$ decays discussed in this chapter:

\beq
 F(z) = \Biggl\{
   \ba{ll}
      1 - \dfrac{4}{z} \arcsin^{2}{(\sqrt{z}/2)}
           \qquad  & z\le 4 \\
      1 + \dfrac{1}{z}\left( \ln\dfrac{1-\sqrt{1-4/z}}{1+ \sqrt{1-4/z}} +
       i\pi \right)^2  & z\ge 4
   \Biggr. \ea
\eeq

\vspace*{.5cm}
\beq
 G(z) = \Biggl\{
   \ba{ll}
      \sqrt{4/z -1} \arcsin{(\sqrt{z}/2)} \qquad & z\le 4 \\
      -{1\over 2}\sqrt{1-4/z}
      \left( \ln\dfrac{1-\sqrt{1-4/z}}{1+\sqrt{1-4/z}} + i\pi \right)
       & z\ge 4
   \Biggr. \ea
\eeq

\vspace*{.5cm}
\beq
\vp(z) = {5\over 18} - {4\over 3 z} - {1\over 3}(1 - {4\over z}) G(z)
\eeq

\vspace*{.5cm}
\beq
H(z) = \dfrac{1}{2(1 - z)^2} \left\{ z F \left( \dfrac{z}{r_\pi^2}\right)
- F \left( \dfrac{1}{r_\pi^2} \right) - 2z \left[ G \left(
\dfrac{z}{r_\pi^2} \right) - G \left( \dfrac{1}{r^2_\pi}\right) \right]
\right\}
\label{eq:H(z)} \eeq
$$
 r_\pi = M_\pi/M_K~.
$$
The kinematical function $\lambda(x,y,z)$ is defined as
\beq
\lambda(x,y,z) = x^2 + y^2 + z^2 - 2 (xy + yz + zx)~.
\eeq

\end{document}